\documentstyle[sprocl]{article}
\input{epsf}
\def\Journal#1#2#3#4{{ #1} {\bf #2}, #3 (#4)}

\def\NPB{{\em Nucl. Phys.} {B}}
\def\PLB{{\em Phys. Lett.} {B}}
\def\PRL{\em Phys. Rev. Lett.}
\def\PRD{{\em Phys. Rev.} {D}}

\def\MPLA{{\em Mod. Phys. Lett.}}
\def\CMP{{\em Comm. Math. Phys.}}

\catcode`\@=11

\def\mycite{\@ifnextchar [{\@tempswatrue\@mycitex}{\@tempswafalse\@mycitex[]}}
\def\mcite{\@ifnextchar [{\@tempswatrue\@mycitex}{\@tempswafalse\@mycitex[]}}

\def\@mycitex[#1]#2{\if@filesw\immediate\write\@auxout{\string\citation{#2}}\fi
 \def\@citea{}\@mycite{\@for\@citeb:=#2\do
    {\@citea\def\@citea{,\penalty\@m\ }\@ifundefined
       {b@\@citeb}{{\bf ?}\@warning
       {Citation `\@citeb' on page \thepage \space undefined}}%
\hbox{\csname b@\@citeb\endcsname}}}{#1}}

\def\@mycite#1{[{#1}]}

\catcode`\@=12

\def\be{\begin{equation}}
\def\ee{\end{equation}}
\def\bea{\begin{eqnarray}}
\def\eea{\end{eqnarray}}

\def\npb#1#2#3{{\em Nucl. Phys.} {\bf B#1} (#2) #3}
\def\plb#1#2#3{{\em Phys. Lett.} {\bf B#1} (#2) #3}

\def\s{\sigma}
\def\g{\gamma}
\def\t{\tau}
\def\a{\alpha}
\def\b{\beta}
\def\d{\delta}
\def\m{\mu}

\def\p{\pi}
\def\e{\epsilon}

\def\l{{\lambda}}

\def\G{\Gamma}
\def\L{\Lambda}
\def\O{\Omega}

\def\apm{\alpha^{\prime}}
\def\CH{{\cal H}}

\def\CM{{\cal M}}

\def\tN{{\tilde N}}

\def\TrH#1{ {\raise -.5em
                      \hbox{$\buildrel {\textstyle  {\rm Tr } }\over
{\scriptscriptstyle \CH _ {#1}}$}~}}
\def\Tr#1{ {\raise -.5em
                      \hbox{$\buildrel {\textstyle  {\rm Tr } }\over
{\scriptscriptstyle{#1}}$}~}}
\def\arrowtop#1{\mathrel{\mathop {\longrightarrow}^{#1 } }  }

\def\IZ{\relax\ifmmode\mathchoice
{\hbox{\cmss Z\kern-.4em Z}}{\hbox{\cmss Z\kern-.4em Z}}
{\lower.9pt\hbox{\cmsss Z\kern-.4em Z}}
{\lower1.2pt\hbox{\cmsss Z\kern-.4em Z}}\else{\cmss Z\kern-.4em Z}\fi}
\def\IC{\relax\hbox{$\inbar\kern-.3em{\rm C}$}}
\def\IR{\relax{\rm I\kern-.18em R}}
\def\1{\relax 1 { \rm \kern-.35em I}}
\font\cmss=cmss10 \font\cmsss=cmss10 at 7pt
\def\viz{{\it viz.}}
\def\ie{{\it i.~e.}}

\def\ad{{\dot a}}
\def\bd{{\dot b}}
\def\cd{{\dot c}}

\def\s{\sigma}
\def\g{\gamma}
\def\t{\tau}
\def\a{\alpha}
\def\b{\beta}
\def\d{\delta}
\def\m{\mu}

\def\p{\pi}
\def\e{\epsilon}

\def\O{\Omega}

\def\l{{\lambda}}

\def\G{\Gamma}
\def\L{\Lambda}

\def\TrH#1{ {\raise -.5em
                      \hbox{$\buildrel {\textstyle  {\rm Tr } }\over
{\scriptscriptstyle \CH _ {#1}}$}~}}

\def\bZ{{\bf Z}}
\def\bR{{\bf R}}
\def\bC{{\bf C}}
\def\bP{{\bf P}}
\def\bS{{\bf S}}
\def\IZ{\relax\ifmmode\mathchoice
{\hbox{\cmss Z\kern-.4em Z}}{\hbox{\cmss Z\kern-.4em Z}}
{\lower.9pt\hbox{\cmsss Z\kern-.4em Z}}
{\lower1.2pt\hbox{\cmsss Z\kern-.4em Z}}\else{\cmss Z\kern-.4em Z}\fi}
\def\IC{\relax\hbox{$\inbar\kern-.3em{\rm C}$}}
\def\IR{\relax{\rm I\kern-.18em R}}
\def\1{\relax 1 { \rm \kern-.35em I}}
\font\cmss=cmss10 \font\cmsss=cmss10 at 7pt

\def\ad{{\dot a}}

\def\frac#1#2{{#1 \over #2}}
\def\ie{{\it i.e.}}

\def\p+{{\partial_+}}

\def\half{{1 \over 2}}

\def\bZ{{\bf Z}}
\def\bM{{\bf M}}
\def\bT{{\bf T}}
\def\tS{{\tilde S}}
\def\tpsi{{\tilde \psi}}
\def\tL{{\tilde L}}
\def\tpsi{{\tilde\psi}}
\def\tq{{\tilde q}}

\def\FL{{(-1)^{F_L}}}
\def\FR{{(-1)^{F_R}}}
\def\ta{{\tilde\alpha}}

\def\apm{\alpha^{\prime}}

\begin{document}
\title{LECTURES ON ORIENTIFOLDS AND DUALITY
\footnote{
Based on lectures given at the 1997 Trieste Summer School on Particle
Physics and Cosmology, Italy. }}
\author{ ATISH DABHOLKAR}
\address{Department of Theoretical Physics\\
School of Natural Sciences\\
Tata Institute of Fundamental Research\\
Homi Bhabha Road,
Mumbai (Bombay), India 400005}
%
%

\maketitle
\abstracts{This is an  introduction
to orientifolds with emphasis on applications
to duality.}

\section{Introduction}

These lecture notes are intended as a pedagogical
introduction to orientifolds.
Aspects of orbifolds and F-theory are also discussed in brief
to provide  the necessary background.
The emphasis is on the  applications
of these constructions to duality.
The approach is based on simple examples
that can be easily worked out in
detail but which, at the same time, 
illustrate
the main ingredients of the general
procedure.  

\subsection{Motivation}

Orientifolds are intrinsically perturbative.
By contrast, much of the recent work in string theory
has focused on explorations of  nonperturbative 
aspects of the theory using the idea of   `duality'.
In view of these developments, it is natural to ask,
before embarking
on the details of the construction,
 why orientifolds are interesting.
Let me begin by addressing this question.

{}From the perspective of duality,
the  motivation
for studying orientifolds is twofold.
{}

1) New dualities:
It  is usually much easier to establish the duality between two theories 
that possess a lot of supersymmetry because, with
more supersymmetry,  the structure of the theory is more 
tightly constrained.
On the other hand, 
theories with less supersymmetry contain more interesting dynamical
phenomena that  are not merely consequences of supersymmetry.
Moreover, to be  closer to the real world,
one would like as little supersymmetry as possible. 
Orientifolds and orbifolds are very useful tools
for establishing  new dualities with less supersymmetry
starting with known dualities with more supersymmetry. 
To illustrate this point, I discuss two  dual pairs, each with
 $16$ supercharges: 

\quad (a) Heterotic string  on a 4-torus  and Type IIA string
on  a $K_3$ surface,

\quad (b)  Heterotic string  on a 2-torus and F-theory
on a $K_3$ surface.\\
We shall see how  these two  dualities can be `derived'
starting with
the $SL(2,\bZ)$ duality of Type-IIB theory which
has $ 32$ supercharges,
and the duality between Type-I string and heterotic string
in ten dimensions.

2) New Compactifications:
The space of string compactifications can have many 
disconnected pieces. Orientifolds  have proved to be
very useful  for exploring different parts of this  moduli space
that were not accessible before as perturbative string vacua.
These new compactifications are often nonperturbatively
connected with known compactifications and have interesting
 duals in M and F theory.

As an illustration I discuss orientifolds in six dimensions with
$8$ supercharges. Many phenomena such as multiple
tensor multiplets or small instantons which appear as
exotic strong coupling effects in the conventional Calabi-Yau
compactifications can be described perturbatively and more explicitly
in the corresponding orientifold duals.
Orientifolds and orbifolds are exact conformal field theories.
Therefore, in principle, one can calculate not only the spectrum but
many other more detailed quantities such as  quantum
corrections to scattering amplitudes by evaluating various
correlation functions in the conformal field theory.
These constructions are thus  complementary to  the geometric
Calabi-Yau compactifications which are simpler to deal with
if one is interested
only in the massless spectrum.

More generally, orientifolds and orbifolds provide us
with `discrete' constructions which may not have
any  geometric interpretation as strings moving 
on some smooth manifold. 
Such non-geometric constructions are 
potentially very interesting
for discovering  new  regions 
of the moduli space which may be disconnected from
known compactifications even at a nonperturbative level.

\subsection{Outline}

My objective will be to give a reasonably self-contained
account of the basic construction of
orientifolds  starting
 with elementary considerations
of free string theory.  After reviewing aspects of 
various related topics, I discuss  some applications to 
duality and compactification. As we shall see,
one can get surprisingly far with this formalism
by keeping track of a few discrete symmetries. 

Our starting point will be the Type-IIB string in
ten space-time dimensions which will be reviewed in  section
$ \S{2}$.
In the sections $ \S{3}$ and $ \S{4}$
the orbifold and orientifold construction will be described by
working
through the examples of 
an orbifold of the Type-IIB string to get Type-IIA string
and an orientifold of Type-IIB string to get
the Type-I string. 
Aspects of the $K3$ surface and Type-II and F-theory 
compactifications on $ K3$ will
be discussed in section $ \S{5}$.
Section $ \S{6}$ deals with applications of orientifolds to duality
of theories with 16 supercharges.
Some orientifolds that give six-dimensional
compactifications with 8 supercharges will be surveyed
in section $ \S{7}$ along with their duals.

\subsection{Orientation}

The remarkable progress in recent years in our understanding of string 
theories has made the subject more exciting and challenging but also 
somewhat less easily accessible for beginners. With the rapidly growing
literature on the subject,
it is not possible to be completely self-contained in this short review.
Many excellent reviews already exist which cover  some of the background
material used here and which 
complement these lecture notes.
I give below a representative but not a very complete list of 
reviews as well as some original articles that can orient the reader.
The two volumes of `Superstring Theory'  by Green, Schwarz,
and Witten \mcite{GSW} 
discuss quantization of free superstring in both GS and
NSR formalism, low energy supergravity 
equations of Type-IIB, Calabi-Yau compactifications along with 
the relevant algebraic geometry. 
A recent review by Sen on duality contains an introduction
to various dualities used here and aspects of nonperturbative string theory.
D-branes are described in detail in the TASI lectures 
 of Polchinski \mcite{PolcTasi}
and in \mcite{PCJ}. More details on orbifolds can be found
in the papers by Dixon, Harvey, Vafa, and Witten~\mcite{DHVW}
and in
the Les Houches lectures of Ginsparg ~\mcite{Gins}.
Various aspects of the $ K3$ surface are discussed in
the review by Aspinwall~\mcite{Aspi}. 
F-theory is introduced and elaborated upon 
in  the papers by Vafa~\mcite{VafaEvid}
and Morrison and Vafa~\mcite{MoVaI,MoVaII}. 

\section{Type-IIB String}

\subsection{Worldsheet Action and Spectrum}

The gauge-fixed, physical spectrum
of the ten-dimensional Type-IIB string 
is easiest to  calculate in the light-cone gauge. 
In the light-cone gauge, the transverse group of rotations 
is $SO(8)$ whose covering group 
is $Spin (8)$.  The three representations
of  $Spin(8)$ that will be  relevant to us
are the vector representation
$\bf 8v$,  
the spinor representation $\bf 8s$,
and the  conjugate spinor representation
$\bf 8c$ which are all eight-dimensional.
The spinor $\bf 8s$ with right-handed chirality is
related to the conjugate spinor $ \bf 8c$ with left-handed
chirality by parity transformation that flips the sign
of one of the components of the vector $ \bf 8v$.
We shall use the letters  $ i, j, k$ as the $\bf 8v$
indices, the letters $ a, b, c$ as the $ \bf 8s$ indices
and the letters, $ \ad, \bd, \cd$ as the $ \bf 8c$
indices. 

In the Green-Schwarz formalism in the light-cone
gauge,
the worldsheet action of the Type-IIB string  is given by~\mcite{GSW}
\bea
\label{eq:action}
S_{l.c.} = {-1 \over 2\pi} \int d\sigma d\tau\,  (\partial_+ X^i \partial_-
X^i  - iS^{ a} \partial_- S^{a}   - i \tilde S^{a}
\ \partial_+ \tilde S^{a}),
\eea 
where $ \sigma$ is the coordinate along
the string, $0\leq \sigma < 2\pi$, and $\tau$ is the worldsheet
time. We have set  $\alpha' = {1 \over 2}$.
In addition to  bosonic fields on the string worldsheet
$X^i$, which are the transverse spatial coordinates of the string, 
there are additional fermionic 
 fields on the worldsheet:
left-moving $ S^a$ and right-moving $\tilde{S}^a$ both 
of which transform as $ \bf 8s$.
Since both right and left movers have the same  spacetime
transformation properties, this theory is non-chiral on the worldsheet.
But since only the right-handed chirality  of spacetime fermions
appears and not the parity transform, the theory is chiral in spacetime.
Another inequivalent choice is to take left-moving $S^\ad$ which
transforms as a  left-handed conjugate spinor and right-moving $\tS^{a}$
which transforms as the right-handed spinor. This choice 
gives Type-IIA theory which has  opposite chirality properties.
To summarize, we have,
\bea
\label{typeii}
\begin{tabular}{lllll}
$S^a$ & $\tilde S^a$ & II B & chiral in spacetime  & nonchiral on world sheet \\
$S^\ad$ & $\tilde S^{{a}}$ & II A & nonchiral in spacetime & 
chiral on worldsheet.\\
\end{tabular}
\eea

Quantization of this $ 1+1$ dimensional free field  theory is
straightforward.
The bosons  $X^i$ satisfy periodic boundary condition along
$\sigma$, and by 
spacetime supersymmetry so do the fermions:
$ S^a (\sigma + 2\pi) =S^a(\sigma )$ etc.
With this boundary condition, the mode expansion is
\bea
\label{mode}
X^i &=& x^i + \half p^i \tau + {i\over 2}\sum_{n\neq 0} 
{1\over n}\,\a_n^i e^{-in (\tau + \sigma)}
+{1\over n}\,  \ta_n^i e^{-in (\tau - \sigma)},\nonumber\\
S^a  &=& {1\over \sqrt{2}}\sum_{-\infty}^{\infty} 
S^a_n \ e^{-in (\tau + \sigma)}, \qquad
\tS^a  = {1\over \sqrt{2}}\sum_{-\infty}^{\infty} 
\tS^a_n \ e^{-in (\tau - \sigma)}.
\eea
Canonical quantization of the fields implies standard
commutation 
and anticommutation relations \mcite{GSW} for the oscillator modes:
\bea
[\alpha^i_m, \a^j_n ] = m \d^{ij}\d_{m+n}, \qquad [\ta^i_m, \ta^j_n ] = m\d^{ij}\d_{m+n}
\nonumber\\
\{S_n^a, S_m^b\} =\d^{ab}\d_{m+n}, \qquad \{\tS_n^a, \tS_m^b\} =\d^{ab}\d_{m+n}.
\eea
The zero modes of the $ X^i$ fields satisfy the Heisenberg commutation relations
$[x^i, p^j] =i\d^{ij},$ and the ground state is therefore labeled by the
momentum eigenvalue $|p\rangle$. 
Note that there are fermionic zero modes as well, $ S^a_0$ and  $\tS^a_0$.
The
ground state should furnish a representation of the 
 zero mode algebra
\bea
\{S^a_0,  S^b_0\} = \delta^{ab}, \qquad \{\tS^a_0,  \tS^b_0\} = \delta^{ab}. 
\label{Clifford} 
\eea
Let us look at the left-movers since right-movers can be treated
similarly. Let us rewrite the anticommutations 
 by defining four fermionic oscillators 
$\sqrt{2} b_m = (S^{2m -1} + i S^{2m}), m =1, \ldots, 4$, which
satisfy the usual anticommutation relations
\bea
\{b_m, b^{\dagger}_n\} =\delta_{mn}, \quad \{b_m, b^{\dagger}_n\}=0, 
\quad \{b^{\dagger}_m, b^{\dagger}_n\}=0.
\eea
This rewriting amounts to choosing a particular
embedding  $SO (8) \supset SU(4) \times U(1)$, 
so that  $\{b_m\}$ transform in the fundamental
representation  $ {\bf 4}$ of $ SU(4)$ with
$ \half$ unit of  $ U(1)$ charge,  which we denote
as $ {\bf 4}(\half)$, and the  $\{b_m^{\dagger}\}$
transform in the complex conjugate representation. 
With this embedding various representations decompose as
\bea
\label{embedone}
{\bf 8v} &=& {\bf 6} (0) + {\bf 1} (1)+ {\bf 1} (-1)\nonumber\\
{\bf 8s} &=& {\bf 4} (\half) + {\bf {\bar 4}} (-\half)\nonumber\\
{\bf 8c} &=& {\bf 4} (- \half) + {\bf {\bar 4}} (\half).
\eea
This embedding is more obvious if we use the fact that
$ SO(6) \sim SU(4)$ and $SO(2) \sim U(1)$.
Then the above is a decomposition of the $SO(8)$ spinor
in terms of the $ SO(6)$ spinor and its conjugate
under the embedding $ SO(8) \supset SO(6) \times SO(2)$.
The representation of Eq.~\ref{Clifford} can
now be worked out easily  by
starting  with the completely `empty' 
 Fock space vacuum $ |0\rangle$ which is
annihilated by  all  $ b_m$'s and then obtaining various
filled states by acting with the creation operators.
One obtains a 16-dimensional representation:
\bea
\label{rep}
\begin{tabular}{llllll}
$ |0\rangle$   & & {\bf 1}(1) && \\
$ b^{\dagger}_m|0\rangle$ && $ {\bf {\bar 4}}(\half$) && \\
$ b^{\dagger}_m b^{\dagger}_n |0\rangle$ && {\bf 6}(0) && \\
$ b^{\dagger}_m b^{\dagger}_n b^{\dagger}_p |0\rangle$ && $ {\bf 4}(-\half)$&& \\
$ b^{\dagger}_mb^{\dagger}_n b^{\dagger}_p b^{\dagger}_q
|0\rangle$ && {\bf 1}(-1) && \\
\end{tabular}
\eea
where the labels in the second column indicate the dimensions of the $ SU(4)$ representation and the $ U(1)$ charges.
We see from the $ SU(4)\times U(1)$ quantum numbers
that  16-dimensional
representation of  
the left-moving ground states reduces as  a  sum of two representations
${\bf 8v} +{\bf 8c}$. Similarly, for the right-movers, the
ground states are given by 
the  sum of ${\bf 8v} +{\bf 8c}$.

A string state $ |\psi\rangle$ is constructed by acting 
with various creation operators on
this $16\times 16$-dimensional ground state 
carrying some spacetime
momentum $p$.
A physical state is subject to
the on-shell  conditions:
\bea
\label{virasoro}
\apm M^2 = -\apm p^{\mu} p_{\m} &=& 4( N \equiv \sum_{n=0}^{\infty} n \a^i_{-n}\a^i_n
+S^a_{-n}S^a_{n})\nonumber\\
&=&4(\tN \equiv \sum_{n=0}^{\infty} n \ta^i_{-n}\ta^i_n
+\tS^a_{-n}\tS^a_{n}).
\eea
We see that the  massless states  have
no oscillator excitations and 
 can therefore be read
off by tensoring the left-moving and right-moving ground states:
\bea
(|i \rangle \oplus | \dot{a} \rangle ) \otimes (|j \rangle \oplus |\dot{b} \rangle).
\eea

In the Neveu-Schwarz-Ramond formalism
of the superstring, which will be reviewed in $ \S{2.4}$, and 
which is equivalent to the Green-Schwarz 
formalism that we have used here, 
the {\bf 8v} comes from the Neveu-Schwarz (NS) sector,
whereas the 
(NS) and the $ {\bf 8c}$ comes from  the Ramond (R) sector. 
The Neveu-Schwarz states are spacetime bosons whereas
Ramond states are spacetime fermions. 
{}In the tensor product of left-moving and right-moving
states,  the NS-R and R-NS sector give rise to spacetime fermions 
$ \psi_{i \ad}$ and $ \psi_{j \bd}$ which are the two gravitini
of Type-IIB string. The NS-NS states $|i\rangle \otimes |j\rangle$
 can be reduced in terms of the symmetric traceless, antisymmetric,
and scalar combinations which give rise to 
the metric  $ g_{ij}$, the 
2-form $ B_{ij}$, and the dilaton $ \phi$, respectively.
The R-R states 
$|\dot{a}\rangle \otimes |\dot{b}\rangle $
can be reduced as
\bea
\label{reduce}
 \lambda^{\dot{a}}_1 \lambda^{\dot{b}}_2 \sim 
\lambda^T_1 \lambda_2 \oplus  \lambda^T_1 \Gamma^{ij} \lambda_2\oplus
\lambda^T_1 \Gamma^{ijkl} \lambda_2,
\eea
in terms of the Gamma matrices $\Gamma^i$, and their totally
antisymmetrized products $\Gamma^{ij}$ and $\Gamma^{ijkl}$.
Because $\lambda_1$ and $\lambda_2$
have the same chirality,
products such as $ \Gamma^{i}$ and $ \Gamma^{ijk}$
do not appear, and moreover, the combination
$ \lambda^T_1 \Gamma^{ijkl} \lambda_2$ is required
to be self-dual. Altogether we  obtain a scalar $ \chi$,
a 2-form $ B'_{ij}$, and a self-dual 4-form $ D_{ijkl}$
from the R-R sector.  In summary, the massless
spectrum of Type-IIB is as follows.

\noindent {\underline {Bosons}}: 

NS-NS:
metric $g_{ij}$, 2-form $ B_{ij}$, {\rm dilaton} $\phi$,

R-R:
scalar $ \chi$, 2-form $ B'_{ij}$, self-dual 4-form $D_{ijkl}$.

\noindent {\underline{ Fermions}}: 

NS-R: gravitino $ \psi_{i\ad}$.

R-NS: gravitino $ \psi_{j\bd}$.

\subsection{Perturbative Symmetries} 

There are two  perturbative $\bZ_2$ symmetries of the Type-IIB 
string which will be of particular interest to us
\footnote{By perturbative symmetry we mean here a symmetry
that is evident at the perturbative level but which is believed
to be unbroken even nonperturbatively. A nonperturbative symmetry
 by contrast is not evident at the perturbative level.}. 

1) $ \Omega$: As we have seen, Type-IIB theory  is non-chiral
on the worldsheet. Hence, worldsheet parity  $ \Omega$
which reverses the  orientation of the string ($ \sigma \rightarrow 2\pi-\sigma$ )
is a  symmetry of  the theory. Orientation reversal takes right-movers
to left-movers. 
Therefore, of the  states
$ |i\rangle \otimes |j\rangle$ coming from the NS-NS sector,
the symmetric combinations are even 
and antisymmetric combinations are odd under $ \Omega$.
 For  the states $ |a\rangle \otimes |b\rangle$ coming from
the R-R sector, we have to remember Fermi statistics
under exchange.
To summarize,

$ g_{ij}, \phi, B'_{ij}$ are even under $ \O$, 

$ \chi, B_{ij}, D_{ijkl}$ are odd under $ \O.$

$ \Omega$ takes the NS-R states to R-NS states,
thus one combination of the two gravitini  is even under $ \O$
and the other is odd.

2)  $\FL$:  The action in  Eq.~\ref{eq:action}
is invariant under $ S^a\rightarrow -S^a$. 
This symmetry can be written as $ \FL$, where $F_L$ is
spacetime fermion number coming from left-movers.
Only  left-moving fermions are odd under this symmetry,
so  R-NS and R-R states are odd 
whereas the NS-R and NS-NS states are even.
To summarize,

$ g_{ij}, \phi, B_{ij}$ are even under $ \FL$, 

$ \chi, B'_{ij}, D_{ijkl}$ are odd under $ \FL.$\\
The two elements  do not commute with each other. 
In particular, 
\bea
\O \FL\O~=~\FR,
\eea where $F_R$ is 
the right-moving spacetime fermion number and 
$ \FR$ 
takes $ \tS^a$ to $ -\tS^a$.
If we consider all distinct products of these elements,
we get an eight-element nonabelian group as the
group of discrete perturbative symmetries of Type-IIB.
This group is isomorphic to $ D_4$-- the group of
symmetries of a square--which is a subgroup of
$ O(2)$, the group of rotations and reflections in 
the $ x-y$ plane.
$ D_4$ is generated by  two elements:
A reflection, $(x, y)\rightarrow (-x, y)$,
which we identify with $ \O$,
and a rotation through $ \pi/2$: $ (x, y) \rightarrow (-y, x)$, which
we identify with $ \Omega\FL$.
The full 
group of perturbative symmetries is
\bea
G =\{1, \O\FL, (-1)^F, \Omega (-1)^{F_R}, \FL, \O (-1)^F, \FR \}.\eea

\subsection{Nonperturbative Symmetries}

Apart from the perturbative symmetries,
the Type-IIB string has an  $SL(2, \bZ)$ 
duality symmetry. Part of this duality symmetry
 is nonperturbative because as we shall see it
relates a theory at strong
coupling to the dual theory at weak coupling.
The great utility of duality,
is that we can  learn about the strong coupling
behavior of a theory from  the weak coupling behavior
of the dual theory \mcite{SenNonp}.

Unlike the usual perturbative symmetries,
which are valid order by order in perturbation theory,
 a nonperturbative quantum symmetry
is not apparent at the perturbative
level. Establishing the existence of a nonperturbative
symmetry, in general, would require the knowledge of
of the full quantum theory including its strong coupling behavior. 
What makes duality 
useful is that, in a supersymmetric theory,  it is
often possible to discover the duality
symmetries only from the weak coupling, semiclassical data
without having to know all the details of the full quantum theory.
Supersymmetry places powerful constraints on the structure of 
the quantum theory.
With enough supersymmetry,
many semiclassical quantities receive no quantum corrections--either
perturbative or nonperturbative-- and are exact. 
One can analyze such quantities in the
weak coupling regime,
and then analytically continuing them in the strong coupling regime.
For 
two theories to be dual to each other, all such 
nonrenormalized quantities  must match.
This requirement provides many nontrivial consistency checks on
the possible duality symmetry.  The verification of these consistency
checks is often  sufficiently
compelling to establish duality even though one cannot actually
`prove' it.

The Type-IIB
theory has  $N=2$ chiral supersymmetry in ten dimensions
with 32 real supercharges. 
This supersymmetry is highly restrictive.
In fact, Type-IIB string is the only string theory,
and the Type-IIB supergravity 
is the
only possible supergravity 
with N=2 supersymmetry in ten dimensions. 
As long as the strong coupling effects do not
break supersymmetry, the only theory that a Type-IIB
theory can possibly be dual to is Type-IIB theory itself.
The Type-IIB theory is indeed  self-dual with duality
group $SL(2, \bZ)$.  
We now indicate some evidence for this claim.

There are two semiclassical quantities that are expected not to get
renormalized which must  exhibit the $SL(2, \bZ)$ symmetry. 
These are,

1) the massless spectrum and their equations of motion,

2) the spectrum of all BPS-saturated supersymmetric states.

\noindent We shall discuss the first point in this subsection, and return
to the second point in subsection $ \S{2.6}$.

The  low energy equations of motion of Type-IIB string
are given by the Type-IIB supergravity \mcite{GSW}.
These equations are indeed invariant under
the action of a noncompact group of 
symmetry $ SL(2, {\bf R})$, which is the group of $ 2\times 2$
real matrices with determinant one.
A general element
of $ SL(2, {\bf R})$  is 
$ \Lambda =\pmatrix{a&b \cr c&d}$ with $ a, b, c, d$ real and
$ ad-bc =1$. To exhibit the action of the symmetry on massless
fields, let us define a complex scalar
$ \lambda = \chi + ie^{-\phi}$ and  Einstein metric
$ G_{MN} = e^{\phi/2}g_{MN} $.
The field $ \l$ parametrizes  
the upper half plane. 
The action of $ SL(2, {\bf R})$ on
the bosonic fields is given by
\bea
\label{sl2r}
\lambda \rightarrow {a\lambda + b \over c \lambda + d}\, , \quad
\pmatrix{B\cr B'}
\rightarrow \pmatrix{d & -c \cr -b& a} \pmatrix{B \cr B'}, \quad D\rightarrow D,
\quad G\rightarrow G.
\label{transform}
\eea

At the quantum level, the full $ SL(2, {\bf R})$ symmetry
does not survive. The reason is that the theory contains 
states that are charged
with respect to the fields  $B$ and $B'$. The charges   
are  integers satisfying
the  Dirac quantization condition and they
transform  linearly in
the same way as the gauge fields $ B$ and $ B'$
that they couple to.
After a  general $SL(2, {\bf R})$ transformation, the
charges would no longer be integers and would not
 respect the quantization condition. 
However, an  $SL (2, \bZ)$ subgroup that
consists of  $ SL(2, {\bf R})$  matrices with $a, b, c, d $ all integers
does not change the integrality of charges. This subgroup
can be, and in fact is,  an exact duality symmetry  of Type-IIB theory.
The  $SL (2, \bZ)$ is generated by the elements:
\bea
\label{generate}
T &:& \lambda \rightarrow \lambda + 1,  \qquad \L=\pmatrix{1&1 \cr 0&1}
\nonumber \\ 
S &:& \lambda \rightarrow -1/\lambda, \qquad \L =\pmatrix{0&1 \cr -1 &0}
\nonumber \\
R &:& \lambda \rightarrow \lambda, \qquad\quad\,\,\,\,\, \L =
\pmatrix{-1&0 \cr 0&-1}.
\end{eqnarray}

\centerline{\underline{Observations}}

\begin{figure}[htb]
\begin{center}
\leavevmode
\hbox{%
\epsfxsize=3.0in
\epsffile{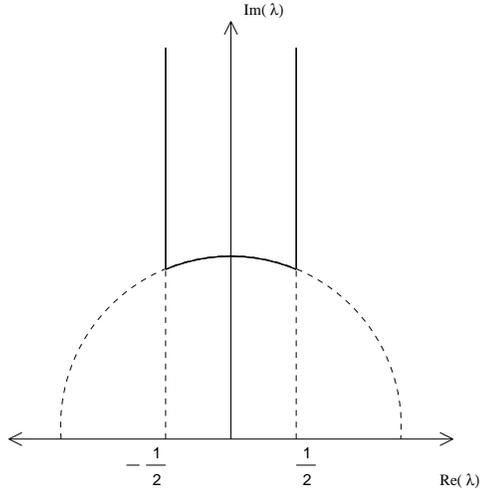}}
\caption{The moduli Space of Type-IIB}
\label{fig:modular}
\end{center}
\end{figure}

1)  A given vacuum of the Type-IIB theory is characterized by
an expectation value of the scalar field $ \lambda$.
$ \l$ parametrizes the upper half-plane,
which can also be written as a coset $ SL(2, \bR )/ SO(2)$.
The duality symmetry $ SL(2, \bZ)$ is a discrete gauge symmetry;
it says that a theory at a given $ \l$ is nonperturbatively equivalent
to all theories at the images of $ \l$ under $ SL(2, \bZ)$.
All gauge equivalent theories must be identified.
The moduli space of Type-IIB is therefore 
$ SL(2, \bZ) \backslash SL(2, \bR )/ SO(2)$ which is as shown in Figure~\ref{fig:modular}.

2) The expectation value of $e^\phi$ is the string coupling constant.
Weak coupling corresponds to $ e^{-\phi}= Im (\lambda )\rightarrow \infty$.
If we set $ \chi =0$ then the element $ S$ takes $e^{\phi}$ to $ e^{-\phi}$,
and thus relates a theory at strong coupling to a theory at weak coupling.

3)  $ R=(-1)^{F_L} \Omega$.
This  can be checked easily from their action
on the massless spectrum. Under
$ R$, the two 2-forms $ B$ and $ B'$ are odd, and all other fields are even,
which we see, 
{}from $ \S 2.2$, is the same action as $ (-1)^{F_L} \Omega$.

4) $S(-1)^{F_L} S^{-1} = \Omega$. 
This can be immediately verified from
 $ \S 2.2$ and Eq.~\ref{sl2r} and 
will be important later when we discuss
$F$-theory. 

\subsection{Neveu-Schwarz-Ramond Formalism}

In the NSR formalism in the light-cone
gauge, the worldsheet fermions $ \psi^i$ and $ \tpsi^i$
transform as a spinor on the worldsheet but as a vector,
$\bf 8v$, of Spin(8).
The worldsheet action of the Type-II string  is given by~\mcite{GSW}
\bea
\label{eq:actiontwo}
S_{l.c.} = {-1 \over 2\pi} \int d\sigma d\tau\,  (\partial_+ X^i \partial_-
X^i  - i\psi^{ i} \partial_- \psi^{i}   - i \tpsi^{i}
\ \partial_+ \tpsi^{i}).
\eea 
The bosons satisfy periodic boundary condition 
and are treated as before.
Fermions can
be either periodic or antiperiodic on the left and on the right.
In each sector one has to perform the GSO projection to obtain
the superstring \mcite{GSW}.

Let us look at the left-movers.
Antiperiodic boundary condition for the fermions gives
the Neveu-Schwarz sector of the theory. The bosons are integer-moded
but the fermions are half-integer moded.
The ground state $ |{\rm NS}\rangle$
has energy 
$ -\half$ and is tachyonic.  A state
with oscillator number $ N$ has mass $ M$ which satisfies
the mass-shell  condition
\bea
\apm M^2 = 4( N-\half).
\eea
The ground state is therefore tachyonic. 

The GSO projection projects out states with odd worldsheet fermion number
$f$.
The ground state is assigned odd worldsheet fermion
number,  $(-1)^f =-1$. This assignment of the fermion number follows
from the fact that the `vacuum'  $ |{\rm NS} \rangle$ actually
has a fermionic ghost excitation \mcite{PolcTasi}. 
It is also the choice that projects
out the tachyon after the  GSO projection.
The GSO-projected spectrum contains a 
 massless state 
\bea
\psi^i_{-\half}|{\rm NS}\rangle,
\eea
which transforms as a vector $ \bf 8v$ of $ Spin(8)$.

Periodic boundary condition for the fermions gives the Ramond
sector. Now, the bosons and fermions are both integer moded,
and the ground state energy is zero.
There are $ 8$ zero modes  $ \psi^i_0$
with anticommutations
\bea
\label{NSanti}
\{ \psi_0^i, \psi_0^j \} =\delta^{ij}.
\eea
With $ \psi^i =\sqrt{2}{\Gamma^i}$, Eq.~\ref{NSanti} defines the usual
Clifford algebra of Dirac matrices. 
The representation of this algebra can be found by
following steps similar to those that led to 
Eq.~\ref{rep} in
$ \S{2.1}$.
Now  the fermionic oscillators are defined
by ${\sqrt{2}}d_m = \psi^{2m -1} + i \psi^{2m}, m =1, \ldots, 4$, which
satisfy the usual anticommutation relations
$$\{d_m, d^{\dagger}_n\} =\delta_{mn}, \quad \{d_m, d^{\dagger}_n\}=0, 
\quad \{d^{\dagger}_m, d^{\dagger}_n\}=0.$$
This definition amounts  to a  different
embedding  $ SO (8) \supset SU(4) \times U(1)$
than Eq.~\ref{embedone}.
Various representations now decompose as
\bea
{\bf 8v} &=& {\bf 4} (\half) + {\bf {\bar 4}} (-\half)\nonumber\\
{\bf 8s} &=& {\bf 4} (- \half) + {\bf {\bar 4}} (\half).\nonumber\\
{\bf 8c} &=& {\bf 6} (0) + {\bf 1} (1)+ {\bf 1} (-1)
\eea
The Fock space of the fermionic oscillators furnishes a 16-dimensional
representation. One can define the chirality matrix
\bea
\label{gamma}
\G = \G^1 \G^2 \ldots \G^8,\qquad (\G)^2=1, \qquad
\{\Gamma, \Gamma^i\} =0.
\eea 
On left handed fermions,  $ \G =1$, and on right-handed fermions
$ \G =-1$.
The 16-dimensional representation  reduces as
${\bf 8s} +{\bf 8c}$. 
The similarity between the the algebra of  $ \psi_0^i$'s and of 
$ S^a_0$'s  is a reflection of the triality symmetry of the Spin(8) algebra  
which  interchanges  the three eight-dimensional
representations $ \bf 8v$, $ \bf 8s$, and  $ \bf 8c$ into each other.
The Clifford algebra and the representations are related by triality:
\bea
\label{gammai}
\begin{tabular}{llll}
\qquad\qquad $\Gamma^i \in {\bf 8v}$, & $\{\Gamma^i, \Gamma^j\} = 2 \delta^{ij}$ & 
gives & ${\bf 8s} \oplus {\bf 8c}$\\
\qquad\qquad $\Gamma^a \in {\bf 8s}$, & $\{\Gamma^a, \Gamma^b\} 
=2 \delta^{ab}$ & gives & ${\bf 8v} \oplus {\bf 8c}$ \\
\qquad\qquad $\Gamma^{\dot{a}} \in {\bf 8c}$, & $\{\Gamma^{\dot{a}}, 
\Gamma^{\dot{b}}\} = 2 \delta^{\dot{a}\dot{b}}$& 
gives & ${\bf 8s} \oplus {\bf 8v}$ 
\end{tabular}\nonumber\\
\eea
Indeed, the triality of the Spin(8) algebra is what makes the 
equivalence between the NSR and the GS formalism possible
\mcite{GSW}.

Similarly, there is a NS and R sector for the right-movers. 
GSO projection in the R sector keeps only one the two spinors. 
The relative choice of the GSO projection
for  the right-movers and for the left -movers is significant:
we  can keep either fermions of the same
chirality or of opposite chirality in the two sectors. 
Depending on the choice, we get either Type-IIA theory
or Type-IIB theory:
\bea
{\rm Type\, IIA}: && ({\bf 8v} \oplus {\bf 8s}) \otimes ({\bf 8v} \oplus {\bf 8c})
\nonumber\\
{\rm Type\, IIB}: && ({\bf 8v} \oplus {\bf 8c}) \otimes ({\bf 8v} \oplus {\bf 8c})
\eea

\subsection{ T-duality}

Consider a single periodic boson $ X$ with period 
$ 2\pi R$, which can be thought of as a coordinate of a string on a
circle of radius $R$.
The momentum along the circle is now quantized, 
$ p =n/R$. Moreover, the string can wind around the circle before closing,
so there are different topological sectors labeled
by the winding number $ w$. In sector with winding number $ w$,
$X$ satisfies the boundary condition,
$X(\s +2\pi, \t ) =X(\s) + 2\pi w R$. The
 mode expansion of $ X$ in each sector
is similar to Eq.\ref{mode}:
\bea
X(\s, \t)  \sim x + {n \apm\over  R} \tau + w R\s +{\rm oscillators}.
\eea
We can write 
 $ X(\s, \t)= X_L(\s^+ ) +X_R(\s^-)$, where $ X_L$ and $ X_R$
are left-moving and right-moving fields respectively
with $ \s^+ =\t +\s$ and $ \s^- =\t-\s$. Their 
mode expansion is given by
\bea
X_L(\s^+ ) = x_L + \sqrt{\apm \over 2} (q \s^+ +
i\sum_{n\neq 0}{1\over n}\, \a_n e^{-in \s^+}),\nonumber\\
X_R(\s^-) = x_R + \sqrt{\apm \over 2}({\tilde q} \s^- +
i\sum_{n\neq 0} {1\over n}\,  \ta_n e^{-in \s^-}),
\eea
where 
\bea\label{momenta}
q&=& \left({n\over R}+{wR\over\apm}\right)
\sqrt{\apm\over2} \nonumber\\
{\tilde q} &=&
\left({n\over R}-{wR\over\apm}\right)\sqrt{\apm\over2}.
\eea

The Hamiltonian for the boson is 
\bea
\label{mass}
H = ({q^2\over 2} + N ) + ({\tq^2\over 2} + \tN ),
\eea
where $ N$ and $ \tN$ are the oscillator numbers.
{}From Eqs.~\ref{momenta} and \ref{mass}, it is easy to check that the
spectrum is invariant under  the `T-duality' transformation,
$ R\rightarrow \apm/R$ and $ n\leftrightarrow w$, which
takes $ q\rightarrow q$ and $\tq\rightarrow-\tq $.
At the level of the field $X$, 
T-duality can be thought
of as a one sided parity transform,
\bea
\label{onesided}
X_L \rightarrow - X_L, \qquad \
X_R \rightarrow X_R,
\eea
taking $ R$ to $ \apm/R$ at the same time.

This symmetry of a free-boson in two dimensions has
a remarkable interpretation in Type II string theory.
It  relates Type-IIB  string  compactified on a circle of  radius $ R$ 
to Type-IIA string  compactified on a circle of radius $ \apm/R$.
In string theory, there are additional fields on the worldsheet.
If we T-dualize along a direction $ 9$, then other bosonic
coordinates are not affected, but the
fermions must transform in accordance with spacetime supersymmetry.
By spacetime supersymmetry, T-duality must act like a left-sided parity
transform even for spacetime fermions:
\begin{eqnarray}
X^9_L\rightarrow -X_L^9,\qquad S^a &\rightarrow&  \Gamma\Gamma^9 S^a \nonumber\\
X^9_R\rightarrow X_R^9,\qquad \tilde S^a &\rightarrow& 
\tilde S^a,
\end{eqnarray}
where the Gamma  matrices are as defined in Eq.~\ref{gammai} 
and Eq.~\ref{gamma}.
$ \Gamma\Gamma^9$  represents the action of the parity 
transformation $X^9\rightarrow X^9$ on the $\Gamma$ matrices,
because it anticommutes with $ \G^9$
 ($\Gamma^9\rightarrow -\Gamma^9$) and commutes with
$ \G^j$ 
($ \Gamma^j\rightarrow \Gamma^j$) for $j\neq 9$.
This operation changes the chirality of right-moving fermions,
because $\Gamma\Gamma^9 S^a$ transforms as a conjugate spinor
with a dotted index $\ad$.
We thus get, on the worldsheet a left-moving conjugate
spinor ($ {\bf 8c}$) and right-moving spinor ($ {\bf 8s}$), 
which  is
Type-IIA string theory.
To summarize, the T-duality $T_9$ takes
 Type IIB on a circle of radius $R $ in the $ X_9$ direction to 
{\rm Type~IIA~on~the dual~circle~of~radius}~$ \apm/R$.

T-duality is a symmetry not only of the free string theory but 
also of the interacting
theory.
Indeed, the worldsheet
path integral of the boson on a higher genus Riemann surface
can be shown to be  invariant under this transformation
\mcite{GPR}. 
Therefore, by factorization, not only the free spectrum, but also
string interactions  respect this symmetry.

Let us see what happens to various symmetries of Type II theory under
T-duality.
\bea
\Omega \ {\rm in~~II~B} \mathrel{\mathop {\longrightarrow}^{T_9 } } 
I_9 \Omega \ {\rm in~~II~A}
\eea
where $ I_9$ is the inversion of the 9-th coordinates:
\bea
I_9:\ \ (X_L^9,  X^9_R) \rightarrow (-X_L^9, -X^9_R).
\eea
In other words,
\bea
T_9 \Omega T_9^{-1} = I_9 \Omega,
\eea
as can be seen  from
\bea
(X_L^9, X^9_R) \arrowtop{T_9^{-1}}(-X^9_L, X_R^9)
\arrowtop{\Omega} (X^9_R, -X^9_L)\arrowtop{T_9} (-X^9_L, -X^9_R).
\eea
On fermions, $ I_9$ is a parity transformation for both left-movers
and right-movers
\bea
I_9: \ \ (S^a, \tS^b) = 
(\Gamma\Gamma^9 S^a, \tilde\Gamma \tilde\Gamma^9 \tS^b),
\eea
which flips the chirality of both fermions.
Note that $\Omega$ by itself is {\it not} a symmetry of  Type-IIA because
starting with a 
left-moving spinor $ S^a$ and right-moving conjugate spinor $ \tS^\ad$
we get a left-moving conjugate spinor $ S^\ad$ and right-moving
spinor $ \tS^a$. To flip the chiralities this operation has to be followed
by the parity transformation $ I_9$ to get a genuine symmetry:
\bea
({\bf 8s}, {\bf 8c}) \mathrel{\mathop {\longrightarrow}^{\O} }  ({\bf 8c}, {\bf 8s})
\mathrel{\mathop {\longrightarrow}^{I_9 } }  ({\bf 8s},  {\bf 8c}).
\eea

\subsection{Solitons and D-branes}

In $ \S{2.3}$ we discussed the $ SL(2, \bZ)$ invariance
of the effective action of massless states. 
Let us now turn to  the spectrum of massive
BPS-saturated states. BPS states are special states in the spectrum that
preserve some of the spacetime supersymmetries. 
The mass of a BPS-state is proportional to its charge, and 
because of supersymmetry
their spectrum is not quantum corrected \mcite{WiOl}.
If $ SL(2, \bZ)$ is to be the duality symmetry, then the spectrum of
BPS states must be invariant under the $ SL(2, \bZ)$.

The spectrum of perturbative BPS-states by itself is certainly not
invariant under $ SL(2, \bZ)$. For example, take a string that 
winds around the $X^9$ direction as in $ \S{2.4}$ 
but carries no momentum along that direction. 
Such a state is a BPS-state and its mass  is 
proportional to  the winding number \mcite{DaHa}. 
The winding number is in fact the  quantized charge of NS-NS
2-form field $ B_{MN}$. This can be checked easily
from a vertex operator calculation.
Now,  the element $S$ of the duality
group (Eq.~\ref{generate})
takes the NS-NS field $B$ to the R-R field $B'$.
Therefore, for duality to hold, we must find BPS-states that 
that are charged with respect to $ B'$. 
But there are no such states in the perturbative spectrum.
This follows from a general fact that the vertex operator
for the R-R fields involves their field strengths and not the
potential
\mcite{PolcTasi}. As a result, all perturbative states couple to
the R-R field strength and not to the potential.
This coupling of R-R forms to perturbative string
states  is analogous to the coupling of a photon to  a neutron.
A neutron has 
a magnetic moment that couples to the field strength but
has no charge that can couple  minimally to the vector potential.

The 2-form $ B'$ can couple to a string or a one-dimensional brane.
In general, a (p+1)-form from the  R-R sector would
couple to an extended soliton which is a p-dimensional membrane
or a p-brane.
Even though there are no such states in the perturbative spectrum
that couple minimally to
the R-R fields, they do exist in the spectrum as nonperturbative solitons.
What is more, these solitons have an amazingly
simple description in terms of  free open strings
with  mixed Dirichlet and Neumann  boundary conditions
\mcite{PolcDbra,PCJ,PolcTasi}. 

Let us recall some facts about open strings.
The mode expansion
for an open string is very similar to Eq.~\ref{mode}, but at
the end points of the string the left-moving wave gets reflected
and turns into a right-moving wave. It can reflect back either in phase
or out of phase with in the incoming wave,
which corresponds to either Neumann or Dirichlet boundary
condition respectively. Let us take the worldsheet coordinate $\sigma$ along
 open string to run from $ 0$ to $ \pi$. 
Then the boundary conditions 
at both ends of the open string is
\bea
{\rm Neumann}&:&\partial_+X =\partial_-X,\nonumber\\ 
{\rm Dirichlet}&:&\partial_+X = - \partial_-X,
\label{ND}  
\eea
at the ends $\sigma = 0, \pi$.
The mode expansion is
\bea
X(\sigma, 0) =x + i\sqrt{\apm\over 2} \sum_{n\neq 0}{\a_n\over n}
(e^{in\s} \pm e^{-in\s}),
\label{openmode}
\eea
where the $+$ sign is for Neumann boundary condition at both ends
(NN sector), and the $ -$ sign is for Dirichlet boundary condition at both ends
(DD sector).
An important difference between these boundary conditions is that
with Dirichlet boundary condition, the zero mode term $p\tau$ 
in Eq.~\ref{mode}
is absent in the mode expansion. This reflects the fact that
the string cannot move in this direction because the end points
are stuck at the position $ x$.

To describe a  p-dimensional soliton,
consider a p-dimensional
hyperplane along the directions
$X^1, \ldots, X^p$.
Take the longitudinal coordinates
$X^\mu,  \mu=0, \ldots, p$ to satisfy  
 NN boundary conditions, and the transverse coordinates
$X^m, m=p+1, \ldots, 9$ to
satisfy  DD boundary conditions.
These boundary conditions break translational invariance.
Open strings are allowed to end on the p-dimensional
hyperplane which can be viewed as a p-brane 
at a location determined by the zero mode of the coordinates
$X^m$.
This configuration, called a Dirichlet p-brane, behaves in every
respect like a BPS soliton. It couples to gravity and the the
R-R (p+1)-form field. Its mass is proportional to the charge
with respect to the RR field. 
The D-brane worldvolume carries
a  $ U(1)$ supersymmetric gauge theory that is obtained
by dimensional reduction of N=1 super Yang-Mills theory in ten
dimensions to $ p+1$ dimensions.
This can be seen from quantization of the superstring subject
to the Dirichlet and Neumann boundary conditions
\mcite{PolcTasi}. The states
$ \psi^\m_{-\half}|{\rm NS}\rangle$ in the NS sector
give the a vector of $ U(1) $ on the worldvolume
and the states $ \psi^m_{-\half}|{\rm NS}\rangle$ are the scalar superpartners 
in the worldvolume. The Ramond sector gives the fermionic
superpartners.

\begin{figure}[htb]
\begin{center}
\leavevmode
\hbox{%
\epsfxsize=1.0in
\epsffile{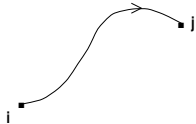}}
\caption{An open string beginning on the $i$-th D-brane and ending on
the $j$-th D-brane.}
\label{fig:ChanPaton}
\end{center}
\end{figure}

If there are  $n$ identical parallel D-branes, then the open string 
can begin on a D-brane labeled by $i$ and end
on one labeled by $j$ (Figure~\ref{fig:ChanPaton}). 
The label of the D-brane is what
in early string theory was called the 
Chan-Paton index at each end. 
Let us denote a  general state in the open string sector
by $|\psi, ij \rangle \l_{ij}$. Here $i, j$ are
Chan-Paton indices, $ \l_{ij}$ is the Chan-Paton wave-function,
$ \psi$ is the state of the worldsheet fields,
and by reality of the string wave function, $\l^{\dagger}=\l$.
The massless excitations of the open string 
now give rise to a supersymmetric $U(n)$ gauge theory
on the worldvolume.

The spectrum of these nonperturbative states provides
many non-trivial checks of duality.  For example, $SL(2,\bZ)$ predicts
a whole tower of  `dyonic' $(p, q)$ strings that 
that have charge $p$ with respect to $B$ and charge $q$ with
respect to $ B'$ \mcite{Schw,WittBoun}. 
Many of these predictions  have now been confirmed
providing  substantial evidence for the correctness of duality.

T-duality has a simple action on D-branes \mcite{PolcTasi}. 
T-duality  along a longitudinal direction 
of a  p-brane turns it into a (p-1)-brane,
and T-duality along a transverse direction
turns it into  a $(p+1)$ brane.
This follows from the observation that T-duality is 
a one-sided parity transform so it turns
Dirichlet and Neumann boundary conditions
into each other.

\section{Orbifolds}

\subsection{General Remarks}

Given a  manifold $ \CM$ with a discrete symmetry $ G$, one can
construct an orbifold $\CM'= \CM /G$. If the symmetry acts
freely on $ \CM$, \ie, without any fixed points, then $ \CM'$
is also a smooth manifold. If there are fixed points then
$ \CM'$ is singular near the fixed points.
If we now consider strings moving on a target space  $ \CM$,
then we are naturally led to the concept of
orbifolds in conformal field theory.

Consider a 
theory $A$ with a discrete
symmetry group $G$.  
One can construct a new theory $A'$ = orbifold of
$A$ by $G$,
$$
A' = A/G .
$$
For simplicity
we  shall take $G$ to be a  $\bZ_2 \equiv \{1, \a\} $ generated by
an involution $\a$ because in fact all examples used in these lectures
are $\bZ_2$ orbifolds.

In point particle theory, we simply take the Hilbert space of $A$ and
keep only those states that are invariant under $G$ to obtain the
Hilbert space of $A'$. However, the particle propagation would
be singular near the fixed points of $ G$.
In closed string theory, we must also add the
``twisted sectors'' that are localized near the fixed points.
In twisted sectors, the string is closed only up to
an action by an element of the group. What is surprising
is that after the inclusion of twisted sectors, string propagation
on the orbifold is nonsingular even near the fixed points.

In string theory, there is a well-defined procedure for
adding   twisted sectors. Twisted sectors are
necessary for modular invariance which is the requirement
that the string path integral be invariant under the modular group.
For a  torus,  the modular group is $SL(2,\bZ )$.
The modular group is the group of global diffeomorphisms   
of the surface. Invariance with respect to this group is essential
to avoid global gravitational anomalies which would render
the theory inconsistent.
This requirement necessitates the inclusion of
twisted sectors. 
We refer the reader
to \mcite{DHVW,Gins} for details of modular invariance.
Physically, {\it unitarity} is what requires twisted sectors.
Even if you excluded twisted states at tree level, once you include
interactions, they will appear in loops because
an untwisted string can split into a string twisted by an element
$ \hat g$ and another string twisted by $ \hat g^{-1}$.

For $\bZ_2$ orbifolds
there are only two sectors: one untwisted and the other twisted 
by $ \a$. In each sector we must perform the projection onto
$ \bZ_2$ invariant states with the projector 
 $ {1\over 2}(1+\hat\a )$.  Here $ \hat \a$ is the operator that 
represents the action of $ \a$ on the Hilbert space.
In the sector twisted by $\a$, all worldsheet fields, which we
generically refer to as $ \Phi$, satisfy the boundary condition
\bea
\Phi(\s +2\pi,\tau) = \hat\a \Phi(\s,\t).
\eea

For $\bZ_2$ orbifolds, level matching is necessary
and sufficient to ensure modular invariance at one loop.
Level matching requires that
\bea
E_L -E_R = 0~{\rm mod}~{\half}
\eea
where $ E_L$ and $ E_R$ are the energies of any two states
of the left-moving and right-moving Hilbert spaces
respectively.
 
In the next subsection I illustrate this procedure by constructing Type IIA theory
as a $ \bZ_2$ orbifold of Type-IIB.

\subsection{Type-IIA theory as an orbifold}

We orbifold Type-IIB theory by the symmetry group
\bea
\bZ_2\equiv\{1, (-1)^{F_L} \}.
\eea

\underbar{Untwisted Sector:}\\
After the projection $ \half( 1 +\FL)$, 
all R-R and R-NS states are removed but the NS-NS and NS-R states
$| |i\rangle  \otimes ( |j\rangle \oplus |\dot{b}\rangle)$ survive.
We are left with
$g_{ij}, B_{ij}, \phi $ and a single gravitino $ \psi{i \bd}$.

\underbar{Twisted Sector:}\\
The twisting of boundary conditions affects only the left-moving
fermion $ S^a$ because other fields are invariant under
$ \FL$.
\bea
\label{twist}
 S^a (\sigma + 2\pi) &=& \FL S^a (\sigma) = -S^a (\sigma)\ \ \ \nonumber \\
\tS^a (\sigma + 2\pi) &=& \tS^a (\sigma), \ \ \ \ \ \ 
X^i (\sigma + 2\pi) = X^i (\sigma). \ \ \ \ \ \ 
\eea
Therefore, the mode expansion of the coordinates $X^i$ and $ \tS^a$
in the twisted sector is the same as in the untwisted sector. 
The oscillators are integer moded as before and, in particular,
the right-moving ground states are given by the representation
of the zero mode algebra
$\{ \tS^a_0, \tS^b_0\} = \delta^{ab}.$
We thus obtain, as in the  untwisted sector,
\bea
|j\rangle \oplus |\dot{b}\rangle
\eea
as the right-moving ground states.

The oscillators of the left-moving fields  are moded with half-integer
modings so as to satisfy the boundary condition \ref{twist}
\bea
S^a  = {1\over \sqrt{2}}\sum_{r=\bZ +\half} 
S^a_r \ e^{-in (\tau + \sigma)}, \qquad
\eea
The ground state energy is a sum of zero point energies of 
the oscillators. For a single complex boson twisted by a 
phase $ e^{2\pi i \eta}$, the ground state energy is given by
the formal sum $\sum_{n=0}^{\infty}\half (n +\eta) $.
It  can be evaluated as $\zeta(0, \eta)$,
where  $\zeta(k, \eta) = \sum_{n=0}^{\infty}\half (n +\eta)^{-k} $ is the  Riemann zeta-function which regularizes the sum.
The ground state energy of a single complex boson  is \mcite{DHVW}
\bea
-{1\over 12} + \half \eta (1-\eta).
\eea
The ground state energy of a fermion with the same twisting
is negative of the above. 
Now, in the left-moving twisted sector,
there are $4$ (complex) bosons which are untwisted $ (\eta =0)$
and and integer moded,
and $4$ complex fermions  that are twisted with  $ (\eta =0)$
and are integer moded.
Adding the zero point energies of these fields we get that
the ground state energy in the twisted sector is $ -\half$.
The ground state is therefore tachyonic because 
The mass-shell condition is
$M^2 =4/\apm (N-\half)$ and 
the level matching condition for physical states is
\bea
N-\half =\tN
\eea
The ground state does not satisfy the physical state condition.
Moreover, it is odd under the action of $\FL$ and is any 
way projected out by the $ \bZ_2$ projection.
The first excited state 
$$
 S^a_{-{1\over 2}} |0\rangle
$$
 satisfies the constraints and the $ \bZ_2$ invariance.
It gives rise to 
 massless states 
\bea
 |a\rangle  \otimes (|j\rangle \oplus |\dot{b}\rangle). 
\eea
We see that an additional gravitino $\psi_{j\a}$ has appeared in the twisted
sector with chirality opposite to the one that was projected out.
The product $|a\rangle \otimes |\bd\rangle$ can be reduced
as in Eq.\ref{reduce}
\bea
 \lambda^{{a}}_1 \lambda^{\dot{b}}_2 \sim 
  \lambda^T_1 \Gamma^{i} \lambda_2\oplus
\lambda^T_1 \Gamma^{ijk} \lambda_2.
\eea
Now, because $\lambda_1$ and $\lambda_2$
have opposite chirality, only
products such as $ \Gamma^{i}$ and $ \Gamma^{ijk}$
appear. We thus obtain a vector  $A_i$ and 
a 3-form $ C_{ijk}$. Altogether
what we have obtained is precisely the spectrum 
of Type-IIA theory, which has two spinors $S^\ad$ and $\tS^b$ to begin
with:

\noindent {\underline {Bosons}}: 

NS-NS:
metric $g_{ij}$, 2-form $ B_{ij}$, {\rm dilaton} $\phi$,

R-R:
vector $A_i$, 3-form $ C_{ijk}$,

\noindent {\underline{ Fermions}}: 

NS-R: gravitino $ \psi_{i\bd}$,

R-NS: gravitino $ \psi_{j a}$.

\section{Type-I String as an Orientifold}

An important and simple example which illustrates most
of the features of the orientifold  construction is Type-I theory
in ten dimensions. In this section we shall work through
this example in detail, after some general remarks about
orientifolds.  

\subsection{General Remarks About Orientifolds}

In general, a symmetry operation of a string theory $ A$ can be a 
combination of target spacetime symmetry and orientation-reversal on the
world sheet. The group of symmetry can then be written as a union
$$
G = G_1 \cup \Omega G_2.
$$
Given such a symmetry of  $A$, one can construct a new theory
$A' = A/G$. In section $ \S{3}$ we had implicitly assumed
that  $ G_2$ is empty and that the orbifold symmetry consists
of only target space symmetries.
If $G_2$ is non-empty, the resulting theory
$A'$ is  called an ``orientifold'' of $ A$
\mcite{DLP,PrSa,BiSa,HoraI,HoraII,Gova,SagnI}.
In most examples discussed recently,  one starts typically with
a ${\bf Z}_N$ orbifold
of toroidally compactified Type IIB theory and then 
orientifolds it further by a symmetry
$\bZ_2 =\{1 , \Omega \b\}$, where 
$\b$ is a ${\bf Z}_2$ involution of the orbifold.
If the orbifold group ${\bf Z}_N$  is generated by the element $\a$,
then the total orientifold symmetry is 
$G= \{1, \a, \ldots, \a^{N-1}, \Omega\b, \O\b\a, \ldots, \O\b\a^{N-1}\}$
or symbolically,
$G=\bZ_N \cup \Omega (\b \bZ_N )$.
We describe below some general features of the orientifold
construction.

 (1) Unoriented Surfaces:\\
An orientifold is obtained, like an orbifold, 
by gauging the symmetry
$G$. A non-empty $\Omega G_2$ means that orientation reversal,
accompanied by an element of $G_2$, is a local gauge symmetry;
a string and its orientation reversed image are gauge equivalent
and must be identified.
Therefore, the  string perturbation
theory  of the orientifold includes unoriented surfaces like
the Klein bottle.

(2) Closed String Sector:\\
The closed string sector of the theory $ A'$ consists of 
states in the Hilbert space of $A$ that are invariant
under $ G$ and which survive the orientifold projection. 
It  is completely analogous to 
the untwisted sector of an orbifold after the projection.
Typically, starting with oriented closed strings, one gets unoriented 
closed strings after the projection.

(3) Tadpole Cancellation and Orientifold Planes:\\
Orientifolds often but not always
 have open strings in addition to the closed strings.
The open string sector in orientifolds is  
analogous to, but not exactly the same as, the twisted sectors  in  orbifolds.
In the case of orbifolds, twisted sectors
are necessitated by the requirement of 
modular invariance.
In the case of orientifolds, the one-loop
diagrams in string perturbation theory include unoriented and open
surfaces for which there is no analog of the modular group.
 There is, however, a consistency requirement for
these surfaces  that is analogous to the requirement of modular
invariance for the torus.
This is the requirement of `tadpole cancellation'.
These loop diagrams  can have a divergence
in the tree channel corresponding to a  tadpole of a massless particle.
Cancellation of all tadpoles is necessary for obtaining a stable 
string vacuum. This requirement 
is very restrictive and it more or less completely
determines when and how the open string should be added.

Physically, nonzero tadpoles imply that the equations of motion
of some massless fields are not satisfied. They occur 
for the following reason.
The planes that are left invariant by the elements of 
$G_2$ are called the `orientifold planes'. Like a D-brane,
an orientifold plane is  a p-dimensional
hyperplane which couples to  an R-R (p+1)-form which we generically
refer to as $ A_{p+1}$.
The charge of the orientifold plane can be calculated by looking the
R-R tadpole, \ie,
 emission of an R-R closed string  state in the zero momentum limit.
If the orientifold plane has a nonzero charge then it acts
as a source term in the equations of motion
for the (p+1)-form field $A_{p+1}$: 
\begin{eqnarray}
dH_{p+2} = \ast J_{7-p} \qquad\qquad d \ast H_{p+2} =\ast J_{p+1},
\eea
where $ H_{p+2}$ is the (p+2)-form field strength of $A_{p+1}$,
$  J_{p+1}$  and $ J_{7-p}$ are  the `electric' and `magnetic' sources.

Consistency of the field equations requires that 
$
 \int _{\Sigma_k}\ast J_{10-k} = 0, 
$
for all surfaces $ \Sigma_k$ without a  boundary.
In particular, there can be no net charge on  a compact space.
This is the analog of Gauss law in electrodynamics.
The field lines emanating from a charge must either
escape to infinity or end on an opposite charge.
In a compact space, the field lines have nowhere to go to 
and hence
must end on an equal and opposite charge.
The only way the negative charge of a p-dimensional orientifold plane
in a compact transverse space can 
be neutralized is by adding the right-number of
Dirichlet p-branes so that Gauss law is satisfied
and all tadpoles cancel.

(4) Open String Sector and Surfaces with Boundaries:\\
 D-branes are hyperplanes where open strings
can end. Inclusion of D-branes
 introduces the open string sector in the theory.
The action of the group $ G$ is represented in the D-brane sector
by some matrices, which we denote by $\gamma$. The $ \g$
matrices act on the Chan-Paton indices:
\begin{eqnarray}
g:\qquad|\psi, ij\rangle \lambda_{ij} \rightarrow |\hat g(\psi), ij\rangle \lambda'_{ij}; \ \
\lambda \rightarrow \l' =\g^{-1}_g \lambda \g_g 
\eea
\bea
\Omega h:\qquad |\psi, ij\rangle \lambda_{ij} \rightarrow 
|\hat{\Omega h}(\psi),  i' j'\rangle \lambda'_{ij}; \ \ 
\lambda \rightarrow \l' = \g^{-1}_{\Omega h} \lambda^T
\g_{\Omega h} 
\label{gaction}
\eea 
Tadpole cancellation together with the requirement
that  the $\gamma$ matrices furnish a  representation of 
 the symmetry $ G$ in the
D-brane sector
determine not only the number
of D-branes but   also  the form
of the $ \gamma$ matrices.  
When $ n$ D-branes coincide, the worldvolume gauge
group is $ U(n)$. After the projection 
 onto $ G$-invariant states, we are left with a  subgroup of $U(n)$.
The group as well as the representations are usually
uniquely determined by the consistency requirements
discussed above.

\subsection{Orientifold Group and  Spectrum of Type-I}

Let me  illustrate  the statements in the previous subsection 
in the context of Type-I theory. Let me first give
the orientifold group and 
the closed and open string spectrum before discussing
tadpole cancellation and consistency conditions.

Type-I theory is  an  orientifold
of Type-IIB theory with orientifold symmetry group 
\bea 
\bZ_2 = \{1, \O\}.
\eea
 
\leftline{\underbar{Closed String Sector:}}
The closed string sector of Type-I theory  contains  unoriented
strings that are invariant under orientation-reversal.
The massless states are simply the
states of Type-IIB that are invariant under $ \O$.
{From} $ \S{2.2}$ we see that only $ g_{ij}, \phi$, $ B'_{ij}$, and
a symmetric combination of the two gravitini survive the projection.

\leftline{\underbar{Open String Sector:}}

Open string sector arises from the addition of  D-branes
that are required to cancel the charge of the orientifold plane.
Orientation reversal is a purely worldsheet symmetry, so it
leaves the entire nine-dimensional space invariant.
Thus, the orientifold plane
is a 9-plane. It turns out to have  $ -32$ units of charge
with respect to the 10-form non-propagating field from the R-R sector.
This charge can be canceled by adding $ 32$ Dirichlet 9-branes
which each have unit charge. The world-volume theory of the D9-branes
gives rise to gauge group $ U(32)$ but only an $ SO(32)$ subgroup
is invariant under the action of $ \O$.

Type-I supergravity super Yang-Mills theory is anomaly free only
if the gauge group is $SO(32)$ or $E_8\times E_8$.
It is satisfying that the spectrum determined by the requiring
worldsheet consistency is automatically anomaly free
\mcite{CaPo,CLNYI,CLNYII}.

\subsection{Loop Channel and Tree Channel}

A massless tadpole leads to  a divergence in tree channel.
For calculating tadpoles it is useful to keep a field theory example
in mind.
Let us consider a very massive  charged particle in field theory
with charge $Q$. 
At low momentum, the charge acts as a
stationary source for a massless photon.
One can calculate the charge $ Q$ of the particle by calculating
the amplitude for vacuum going into a single photon in
the background of this charge.
(Figure~\ref{fig:tadpole})
\begin{figure}[htb]
\begin{center}
\leavevmode
\hbox{%
\epsfxsize=3.0in
\epsffile{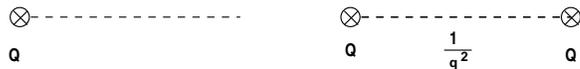}}
\caption{A massless tadpole leads to a divergence in tree channel}
\label{fig:tadpole}
\end{center}
\end{figure}
Alternatively, one can calculate the interaction between two particles
each of charge $Q$ at zero momentum exchange.  
The Feynman diagram has $ 1/q^2$ where
$ q$ is momentum exchange and the residue is proportional
to $ Q^2$.
If we write $ 1/q^2$ as $ \int_0^{\infty} dl \exp{(-q^2l)}$, then
the zero momentum divergence corresponds to the divergence
of this integral  coming from very long
propagation times $ l$.

 D-branes and orientifold planes can be treated similarly.
A D-brane is like a very massive charged particle.
The interaction between the $ i$-th D-brane and the
$ j$-th D-brane due to closed
string exchanges between the two branes
can be computed by evaluating a cylinder diagram with one
boundary on the $ i$-th brane and the other boundary on the
$ j$-th brane.
In string theory, unlike in particle theory,
 because of conformal invariance the tree channel
and loop channel diagrams are related. For example, 
as shown in Figure~\ref{fig:cylinder},
the tree channel cylinder diagram 
can also be viewed
as a loop-channel diagram that evaluates the loop 
of an open string with one end stuck at the  $ i$-th
brane and the other end at the $ j$-th brane.
Similarly, the interaction between an orientifold plane 
and the i-th D-brane is given by the M\"obius strip
diagram which has one boundary stuck at the $ i$-th 
brane and one crosscap stuck at the orientifold plane.
Recall that a crosscap is  a circular boundary with opposite points
on  the boundary identified.  Because some of the elements of the orientifold
group  leave the orientifold plane invariant, the closed
string that emanates from the plane has further identifications
under the symmetry and it  looks like a crosscap.

In summary, we can imagine that  a crosscap is stuck at the orientifold plane 
and the boundary  is  stuck at a D-brane. With an orientifold
with charge $ Q$ and with $ N$ D-branes of unit charge,
the total charge is $ (Q+N)^2$, which can be written as
$ Q^2 + N^2 + 2QN$. The term $ N^2$ is proportional to  the
interaction between the D-branes and  is computed by the cylinder diagram,
the interaction $ 2 Q N$ between the D-branes and orientifold planes
is computed by the M\"obius strip diagram and the interaction
between orientifold planes $ Q^2$ is computed by the Klein bottle diagram.
An efficient way to evaluate these diagrams is to compute them 
in loop channel and then factorize them in tree channel.

The loop-counting parameter in string theory is the Euler character.
A $ k$-th order term in string perturbation theory  which goes
as the $ k$-th power of the string coupling constant $ \lambda$
corresponds
to Riemann surfaces
\begin{figure}[htb]
\begin{center}
\leavevmode
\hbox{%
\epsfxsize=1.0in
\epsffile{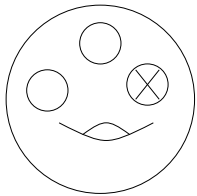}}
\caption{A Surface with two boundaries, one crosscap and one handle}
\label{fig:surface}
\end{center}
\end{figure}
with Euler character $ k-1$.
The Euler character of a Riemann surface with $ b$ boundaries,
$ c$  crosscaps,  and  $ h$ handles is given by 
\bea
\chi = 2 - 2h - b -c.
\eea
A surface with no crosscaps is orientable, otherwise it is nonorientable.
We are  interested in the first  quantum correction, {\it i.~e.}, Riemann
surfaces with $ \chi =0$.
There are four surfaces that contribute:
a torus (one handle), a Klein Bottle (two crosscaps), a M\"obius strip
(one boundary, one crosscap), and a cylinder (two boundaries) (
Figure~\ref{fig:surfaces}).
\begin{figure}[htb]
\begin{center}
\leavevmode
\hbox{%
\epsfxsize=3.8in
\epsffile{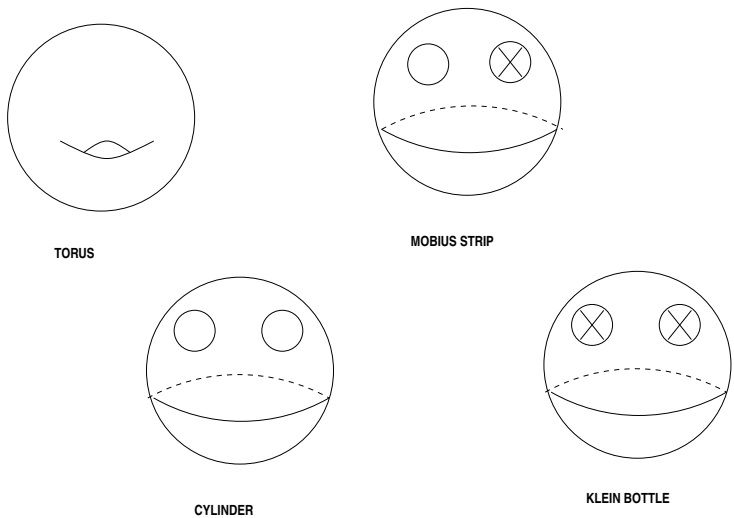}}
\caption{Surfaces with $\chi=0$ in the orientifold perturbation theory}
\label{fig:surfaces}
\end{center}
\end{figure}
Let $ \s_1$ and $ \s_2$ be the coordinate of the surface. Then
we may time slice along constant $ \s_1$ or along constant $ \s_2$.
For a tours, both two time-slicings give a loop diagram,
but for the 
other three surfaces one time slicing give a loop diagram and 
the other time slicing
gives  a tree diagram.  For these three surfaces,
we would like to determine, for later use, what a closed string of 
length $2\pi$ propagating for time $ 2\pi l$ in the tree channel
corresponds to in the loop channel.

The simplest surface  is the cylinder.
Consider, as shown on the left in Figure~\ref{fig:cylinder},
a closed string of length $ 2\pi$ in tree channel, propagating
between two  D-branes  for a 
Euclidean time $ 2\pi l$.
Time runs sideways in the diagram. Now, use the conformal invariance
of string theory to conformally rescale
the coordinates by $ {1\over 2l}$ and take time to run upwards
to get the diagram on the right in 
Figure~\ref{fig:cylinder}.
\begin{figure}[t]
\begin{center}
\leavevmode
\hbox{%
\epsfxsize=1.8in
\epsffile{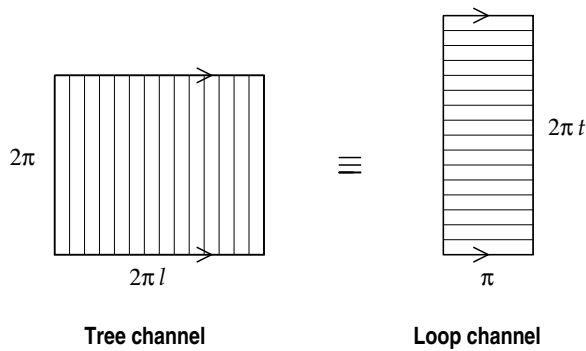}}
\caption{Two ways to view  a cylinder.}
\label{fig:cylinder}
\end{center}
\end{figure}
This diagram represents an open string of length
$ \pi$ beginning on one D-brane and ending
on the other D-brane, propagating in  a loop for a time $ 2\pi t \equiv \pi/l$.
We conclude that $ t = 1/2l$ for the cylinder.

For the  Klein bottle, consider the double cover of the
bottle, \viz, a torus with coordinates $0\leq \s_1\leq 4\pi l$
and $ 0\leq \s_2\leq 2\pi$, with the identifications
$\s_1 \sim \s_1 + 4 \pi l$ and $ \s_2 \sim \s_2 + 2\pi$.
The Klein bottle is obtained by a $ \bZ_2$ identification of the torus:
\bea
\label{eq:bottle}
(\sigma_1, \sigma_2) \sim (4\pi l-\sigma_1, \sigma_2 + \pi).
\eea
We can choose two different fundamental regions.
If we choose the fundamental region as on the top of 
Figure~\ref{fig:bottle},
\begin{figure}[htb]
\begin{center}
\leavevmode
\hbox{%
\epsfxsize=3.6in
\epsffile{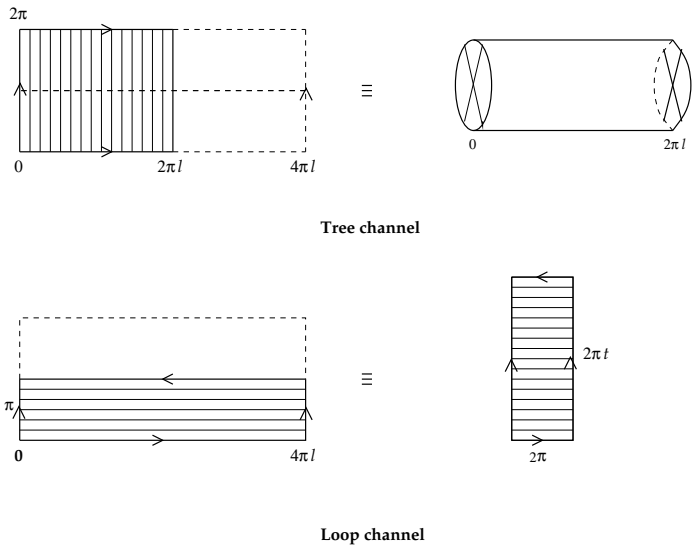}}
\caption{Two ways to view  the Klein Bottle.}
\label{fig:bottle}
\end{center}
\end{figure}
then we get the tree channel
diagram. It represents a closed string propagating between
two orientifold planes for time $ 2\pi l$. If we choose
the fundamental domain 
taking time to run upwards now, as on the bottom of Figure~\ref{fig:bottle}, 
then we have a closed
string propagating in a loop  and undergoing 
a twist $\Omega$. We have to rescale by $ 1/2l$ to obtain a closed
string of length $ 2\pi $ in the loop, which  gives
$ t ={1/4l}$. 

Similarly, for the M\"obius strip we consider the double cover, \viz, 
a cylinder with coordinates $0 \leq \s_1\leq 4\pi l$
and $ 0\leq \s_2\leq 2\pi$, with the identification
$\s_1 \sim \s_1 + 4\pi l$.
The M\"obius strip  is obtained by the  same $ \bZ_2$ identification
as in Eq.~\ref{eq:bottle}.
Again, we  can choose two different fundamental regions.
The fundamental region as on the top of 
Figure~\ref{fig:mobius} gives the tree channel
diagram which represents a closed string propagating between
an orientifold plane and a D-brane.
The fundamental domain as on the bottom of Figure~\ref{fig:mobius},
\begin{figure}[htb]
\begin{center}
\leavevmode
\hbox{%
\epsfxsize=3.6in
\epsffile{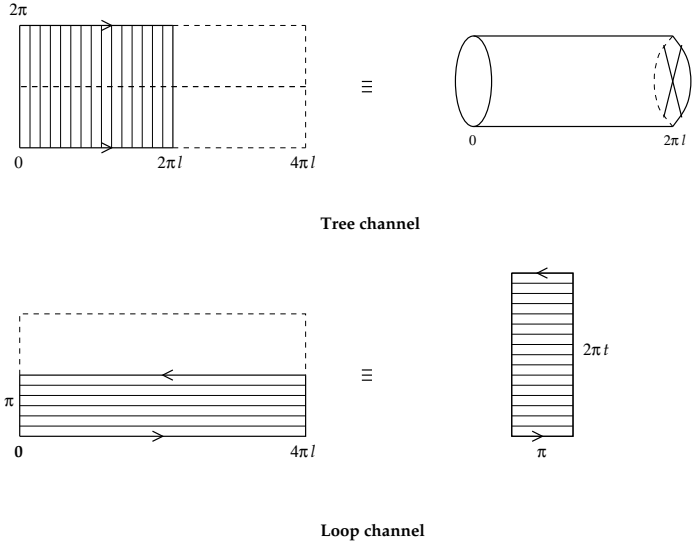}}
\caption{Two ways to view  a M\"obius strip.}
\label{fig:mobius}
\end{center}
\end{figure}
taking time to run upwards now, represents an open string,
with both ends on the  D-brane, 
propagating in a loop  and undergoing 
a twist $\Omega$. We have to rescale by $ 1/4l$ to obtain an open
string of length $ \pi $ in the loop, which gives
$ t ={1/8l}$.

To summarize,  a closed string of length $ 2\pi$
propagating for time $ 2\pi l$ in the tree channel  corresponds 
to an open string of length $ \pi$, or a closed string of length $ 2\pi$
propagating for time $ 2\pi t$ in
the loop channel. For fixed $l$ in the tree channel,
the loop channel time $ t$ for different surfaces is given by
\bea
\label{treeloop}
{\rm Cylinder:}&& t ={1\over 2l}\nonumber\\
{\rm Klein Bottle:}&& t ={1\over 4l}\nonumber\\
{\rm Mobius Strip:}&& t= {1\over 8l} \, .
\eea

We also need to know how the boundary conditions in tree channel
map onto boundary conditions in loop channel. 
In the tree channel,
( $0 \leq \sigma^1 \leq 2\pi l$, $0 \leq \sigma^2 \leq 2\pi$) the
periodicity and boundary conditions on a generic world-sheet
field $\phi$ in the $ g$-twisted sector (see Figure~\ref{treechan})
\begin{figure}
\begin{center}
\leavevmode
\hbox{%
\epsfxsize=2.5in
\epsffile{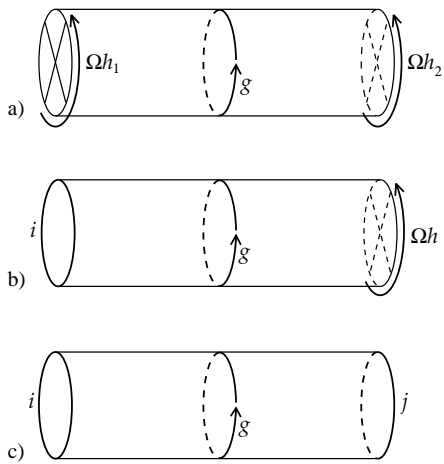}}
\end{center}
\caption{ a) Klein bottle.  b) M\"obius strip. c) Cylinder.}
\label{treechan}
\end{figure}
are as follows:
\begin{eqnarray}
\mbox{KB:}&& \phi(0,\pi+\sigma^2) = \Omega \tilde h_1 \phi(0,\sigma^2),
\quad \phi(2\pi l,\pi+\sigma^2) = \Omega\tilde h_2 \phi(2\pi
l,\sigma^2)\nonumber\\
&& \phi(\sigma^1,2\pi + \sigma^2) = \tilde
g\phi(\sigma^1,\sigma^2)\nonumber\\[4pt]
\mbox{MS:}&& \phi(0,\sigma^2) \in \tilde M_i,
\quad \phi(2\pi l,\pi+\sigma^2) = \Omega  \tilde h \phi(2\pi
l,\sigma^2)\nonumber\\ && \phi(\sigma^1,2\pi + \sigma^2) = \tilde
g\phi(\sigma^1,\sigma^2)\nonumber\\[4pt]
\mbox{C:}&& \phi(0,\sigma^2) \in \tilde M_i,
\quad \phi(2\pi,\sigma^2) \in \tilde M_j,\nonumber\\
&& \phi(\sigma^1,2\pi + \sigma^2) = \tilde g\phi(\sigma^1,\sigma^2). 
\label{treecond}
\end{eqnarray}
Here $ M_i$ is the submanifold where the $ i$-th D-brane is located.
The tilde on the group elements  allows  for  additional
$ \pm$ signs that depend on
 the GSO projection to accompany the action of the group element
 for world-sheet fermions. 
The definitions 
in Eq.~\ref{treecond}
are consistent only if
\begin{eqnarray}
\mbox{KB:}&& (\Omega \tilde h_1)^2 = (\Omega \tilde h_2)^2 = \tilde g
\nonumber\\ 
\mbox{MS:}&& (\Omega \tilde h)^2 = \tilde g, \quad \tilde g \tilde
M_i = \tilde M_i
\nonumber\\
\mbox{C:}&& \tilde g \tilde M_i = \tilde M_i, \quad \tilde g \tilde M_j =
\tilde M_j\ ;
\end{eqnarray}
otherwise the corresponding path integral vanishes.

The loop channel for the Klein bottle and the M\"obius strip
($0 \leq \sigma^1 \leq 4\pi l$, $0 \leq \sigma^2 \leq \pi$) 
is obtained geometrically  by taking the upper strip $\pi \leq \sigma^2 \leq 2\pi$,
inverting it from right to left, multiplying the fields by 
$(\Omega \tilde h_2)^{-1}$, and gluing it to the right side of the lower strip. 
This construction ensures that the fields are smooth at $\sigma^1 = 2\pi l$.
The periodicity conditions are 
\begin{eqnarray}
\label{treeperiod}
\mbox{KB:}&& \phi(\sigma^1,\pi+\sigma^2) = \Omega \tilde h_2 \phi(4\pi l
-\sigma^1,\sigma^2), \quad \phi(4\pi l,\sigma^2) = \tilde g'\phi(0,\sigma^2)
\nonumber\\
\mbox{MS:}&& \phi(\sigma^1,\pi+\sigma^2) = \Omega \tilde h \phi(4\pi l
-\sigma^1,\sigma^2), \quad \phi(0,\sigma^2) \in \tilde M_i,\nonumber\\
&& \qquad\qquad \phi(4\pi l,\sigma^2) \in \tilde M_i
\end{eqnarray}
where $\tilde g' = \Omega \tilde h_2 (\Omega \tilde h_1)^{-1}$.
Rescaling the coordinates to standard length for string in loops
the respective amplitudes are
\begin{eqnarray}
\mbox{KB:}&& \Tr{{\rm closed},g'} \left(\Omega \tilde h_2 (-1)^{f + \tilde
f} e^{\pi (L_0 + \tilde L_0)/2l} \right)
\nonumber\\
\mbox{MS:}&&\Tr{{\rm open}, ii} \left(\Omega \tilde h (-1)^{f}
e^{\pi L_0 /4l}\right)
\nonumber\\
\mbox{C:}&& \Tr{{\rm open}, ij} \left(\tilde g (-1)^{f}
e^{\pi L_0 /l}\right) \label{traces},
\end{eqnarray}
where the closed string trace is labeled by the spacelike twist $g'$ and the
open string traces are labeled by the Chan-Paton labels.

\subsection{Tadpole Calculation}

We have followed  in these lectures
 the formalism and notations  of Gimon and Polchinski
\mcite{GiPo}.
A similar formalism was used for bosonic orientifolds by Pradisi and Sagnotti
in earlier work \mcite{PrSa}.
The tadpole constraints that we are about to describe were applied
to orientifolds also in refs.~\mcite{HoraI,IsOn,BeDu,BiSa,SagnII}. 
An equivalent but technically  different method for calculating
tadpoles is to construct the boundary state and the crosscap state.
We do not use the boundary state method in these lectures but
the details can be found in \mcite{CLNYI,CLNYII,CaPo,Li}.

One-loop amplitude calculates
the one loop cosmological constant in spacetime as  the  sum of zero point
energies of all the fields in the spectrum of the string.
Let us look, for example,  the sum
of zero point energies of the fields in the open string sector:
\bea
\sum_{\rm bosons} {\hbar \omega_{\vec{p}}\over 2} -\sum_{ \rm fermions}
{\hbar \omega_{\vec{p}} \over 2} = -\sum_{i} 
{V_{10}\over (2\pi)^{10}}\int d^{10} p\,
{1 \over 2}
\log (p^2 + m^2_i) (-1)^{F_i},
\label{cosmo}
\eea
where $ m_i$ is the mass and $ F_i$ the spacetime fermion
number of a state $ i$.
Now we use  the identity 
\bea
\log{A}=- \lim_{\e\rightarrow 0} {d\over d\e} A^{-\e} = - \lim_{\e\rightarrow 0}
{d\over d\e}( \e \int {dt\over t^{1-\e}} e^{-2\pi A t}) = -\int {dt\over t} e^{-2\pi A t},
\eea
and
\bea
\apm (p^2 + m^2_i )=  L_0 =
  {\apm p^2 } + \sum_{i=1}^8 \alpha^i_{-n} \alpha^i_{-n} 
+ \sum_{i=1}^8 \psi^i_{-r} \psi^i_r +a,
\eea
where we have included, for convenience,
the normal ordering constant $ a$ in the definition
of $ L_0$.
The sum in Eq.~\ref{cosmo} then equals
\bea
 \int^\infty_0 {dt\over 2t} \, \Tr{\rm NS-R} \exp^{-2\pi t L_o }
\left(1 + {(-1)^f\over 2}\right).
\label{cosmotwo}
\eea
Here $f$ is the worldsheet fermion number,
$\frac{1+(-1)^{f}}{2}$ performs the GSO projection, and the combination
NS$-$R for the trace takes into $ (-1)^{F_i} $ in Eq.~\ref{cosmo}.
The Trace includes the momentum integration
$
{V_{10}\over (2\pi)^{10}}\int d^{10} p,
$
where $ V_{10}$ is, as usual, the regularized volume of a 10-torus that is
taken to be very large to get the theory in 10 flat spacetime dimensions.
 
The trace  Eq.~\ref{cosmotwo} in the canonical formalism equals 
the path integral on a cylinder by the usual relation between
the canonical formalism and the path integral formalism. 
An open string propagates in a loop for Euclidean time $ 2\pi t$ with  time
evolution operator $ \exp^{-2\pi t L_o}$ giving a cylinder diagram.

To obtain the orientifold we have to project onto states
that are invariant under $\O$, which is achieved by
inserting the projector $ {1 +\O } \over 2$ in Eq.~\ref{cosmotwo}. 
The term proportional to $1/2$ corresponds to the cylinder and the term
proportional to $ \O /2$ corresponds to the M\"obius strip.

The resulting partition sums for the Klein bottle,
the M\"obius strip, and the cylinder  are respectively
$\int_0^{\infty}dt/2t$ times
\bea
\label{amplitudes}
{\rm KB:} &\quad & \Tr{\rm NSNS + RR}
\left\{ \frac{\Omega }{2}\,\frac{1+(-1)^{f}}{2}
e^{-2\pi t(L_0 + \tilde L_0)}\right\} \nonumber\\
{\rm MS:} &\quad &\Tr {\rm NS-R}
\left\{ \frac{\Omega }{2}\,\frac{1+(-1)^{f}}{2}
e^{-2\pi t L_0}\right\} \nonumber\\
{\rm C:} &\quad & \Tr{\rm NS-R}
\left\{ \frac{1}{2}\,\frac{1+(-1)^{f}}{2}
e^{-2\pi t L_0}\right\}.
\eea
For the Klein bottle, the sectors NS-R and R-NS are mapped into each
other by $ \O$ and therefore do  not contribute to the trace.
In the closed string sector,
the Virasoro generators $L_0$ and $ \tL_0$  are
\bea
L_0 &=&  {\apm p^2 \over 4} + \sum_{i=1}^8 \alpha^i_{-n} \alpha^i_{-n}  
+ \sum_{i=1}^8 \psi^i_{-r} \psi^i_r + a, \nonumber \\
{\tilde{L}}_0 &=& 
{\apm p^2 \over 4} + \sum_{i=1}^8 {\tilde{\alpha}}^i_{-n} {\tilde{\alpha}}^i_{-n}
 + \sum_{i=1}^8 \tpsi^i_{-r} \tpsi^i_r + {\tilde a},
\eea
where n is summed over integers; $ r$ is summed over integers
in the Ramond sector and over half-integers in
the Neveu-Schwarz sector. The normal ordering constant 
$a$ is $ 0$ in the R sector,
$ -1/2$ in the NS sector, and similarly for $\tilde a$.

The action of $\Omega$ on the modes of the closed string
is
\begin{equation}
\Omega \alpha_r \Omega^{-1} = \tilde \alpha_r, \qquad
\Omega \psi_r \Omega^{-1} = \tilde \psi_r, \qquad
\Omega \tilde\psi_r \Omega^{-1} = - \psi_r
\end{equation}
for integer and half-integer $r$.  The minus sign included in the last
equation  gives the convenient result $\Omega \psi_M \tilde\psi_M
\Omega^{-1} = \psi_M \tilde\psi_M$ for any product $\psi_M$ of mode
operators. If this sign can be omitted it  just corresponds to
$\Omega \to (-1)^f \Omega$, which has the same action on physical states.
In open string, the mode expansions for a  boson
$ X$ are as in Eq.~\ref{openmode}.
Orientation reversal, $X(\sigma, 0) \to
X(\pi-\sigma, 0)$, takes
\begin{equation}
\alpha_m \to \pm e^{i\pi m} \alpha_{m}.
\end{equation}
with the upper sign for NN boundaries conditions and lower for DD.
For fermions, the mode expansions are
\begin{equation}
\psi(\sigma,0) = \sum_r e^{ir\sigma} \psi_r, \qquad  
\tilde\psi(\sigma,0) = \sum_r e^{-ir\sigma} \psi_r.
\end{equation}
Orientation reversal, $\psi(\sigma,0) \to \pm\tilde\psi(\pi-\sigma,0)$, takes
\begin{equation}
\psi_r \to \pm e^{i\pi r} \psi_{r}
\end{equation}
for integer and half-integer $r$.
As for the  closed string there is some physically irrelevant sign freedom.
Following Gimon and Polchinski \mcite{GiPo} we define
\bea
f_{1}(q) = q^{1/12} \prod_{n=1}^\infty \left(1-q^{2n}\right),\ 
&  &f_{2}(q) = q^{1/12} \sqrt{2}\,\prod_{n=1}^\infty \left(1+q^{2n}\right)\nonumber\\
f_{3}(q) = q^{-1/24} \prod_{n=1}^\infty \left(1+q^{2n-1}\right),\ 
&  &f_{4}(q) = q^{-1/24}\prod_{n=1}^\infty \left(1-q^{2n-1}\right),
\eea
which satisfy the Jacobi identity
\bea
{f_{3}^{8}(q) = f_{2}^{8}(q)+f_{4}^{8}(q)}
\eea
and have the modular transformations
\bea
\label{eq:modular}
f_{1}(e^{-{\pi}/{s}}) = \sqrt{s}\,f_{1}(e^{-{\pi} s}),\
f_{3}(e^{-{\pi}/{s}}) = f_{3}(e^{-\pi s}),\
f_{2}(e^{-{\pi}/{s}}) = f_{4}(e^{-\pi s}).
\eea
These combinations are so defined that the normal ordering constant
are automatically taken into account.
The relevant amplitudes are then given by
$
(1-1)\frac{v_{10}}{256} \int_0^\infty \frac{dt}{t^6}
$
times
\bea
\label{sums}
{\rm KB:} &&\quad
 32 \frac{f_{4}^{8}(e^{-2\pi t})}{f_{1}^{8}(e^{-2\pi t})}\cr
{\rm MS:} &&\quad - \frac{ f_{2}^{8}(e^{-2\pi t}) f_{4}^{8}(e^{-2\pi t})}
{f_{1}^{8}(e^{-2\pi t}) f_{3}^{8}(e^{-2\pi t})}
\left\{ {\rm Tr}(\gamma_{\Omega}^{-1}\gamma_{\Omega}^T)
\right\}\nonumber\\
{\rm C:}&&\quad\frac{f_{4}^{8}(e^{-\pi t})}{f_{1}^{8}(e^{-\pi t})}
\left\{ ({\rm Tr}(\gamma_{1}))^2
 \right\}.\eea
We have defined $v_{10} = V_{10} /(4\pi^2 \alpha')^3$ where $V_{10}$ is the
regulated spacetime. The factor $ (1-1)$ corresponds to
NSNS$-$RR exchange. 

The total amplitude in Eq.~\ref{sums} is zero by supersymmetry. 
In the loop channel,
the vanishing is 
 because of the cancellation between the bosonic and fermionic
zero point energies. In the tree channel, there is a different
interpretation. The amplitude vanishes because the graviton-dilaton
exchange is attractive but the RR exchange is repulsive so the net
force is zero. We are of course interested in making sure that the tadpole
of the RR field by iteself is zero.

Let me indicate where all the factors come from.
Consider the cylinder amplitude. Exchange of R-R field 
in the tree channel means
a periodic boundary condition for the worldsheet fermions in
the NSR formalism.  In the loop channel this corresponds
to periodic boundary condition in the Euclidean time direction
which calculates $\Tr{\rm NS-R}(-1)^f$. In the Ramond sector
 $Tr (-1)^f = 0$ because the ground state has equal number
of states that have odd and even fermion number. In the
NS sector the fermions are half-integer moded and the bosons
are integer moded. Therefore,
\bea
\Tr{NS} e^{-2\pi tN} (-1)^f \sim {\Pi (1 - e^{-2\pi t(n+1/2)})^8 \over \Pi
(1 - e^{-2\pi tn})^8}
\sim {f^8_4 (e^{-\pi t}) \over f^8_1 (e^{-\pi t})}
\eea
The momentum integration gives a factor of $ (1/2\apm)^5$ and there
is a factor of $ 1/16$ because of all the $ 1/2$'s in the projectors
in Eq.~\ref{amplitudes}.

The M\"obius strip and the Klein bottle amplitude can be computed
similarly. It is important to keep track of the boundary conditions
for fermions as we go from the loop channel to tree channel
(Eq.~\ref{treeperiod}). 

To factorize in tree channel, we use the modular transformations
Eq.~\ref{eq:modular}
and the Poisson resummation formula
\bea
\sum_{n=-\infty}^{\infty} e^{-\pi (n-b)^2/a} =
\sqrt{a} \sum_{s=-\infty}^{\infty} e^{-\pi a s^2 + 2\pi i s b}.
\eea
Using
the relations Eq.~\ref{treeloop} between $ t$ and $ l$,
the total amplitude for large $l$
becomes
\bea
\label{eq:tadpole}
(1-1) \frac{v_{10}}{16}\int_0^\infty dl
 \left\{ 32^2 - 64 {\rm
Tr}(\gamma_{\Omega}^{-1}\gamma_{\Omega}^T) + ({\rm Tr}(\gamma_{1}))^2
\right\}.
\end{eqnarray}

\subsection{Determination of the Gauge Group}

When we gauge a symmetry group $ G$,
we identify field configurations that are gauge-equivalent.
To be able to to gauge a symmetry, the group
must  furnish a proper representation,
and not merely a projective ({\it i. e.}  representation up to a phase)
representation of a group.
In the open string sector, this requirement places restrictions
on the $ \gamma$ matrices.  For example, in our case,
$\O^2$  should equal  $\bf 1$.
\bea
\Omega^2:\qquad |\psi, ij\rangle \lambda_{ij} \rightarrow 
|\psi,  i' j'\rangle \lambda'_{ij}; \ \ 
\lambda \rightarrow \l' = (\g^{-1}_{\Omega} \g^T_{\Omega}) \lambda(\g^{-1}_{\Omega} \ \g^T_{\Omega})^{-1} 
\eea
implying that
\begin{eqnarray}
\g^{-1}_{\Omega } \ \g^T_{\Omega} &=& {\bf 1},
\end{eqnarray}
or
\begin{equation}
\gamma_{\Omega}^T = \pm \gamma_{\Omega}. 
\label{chitran}
\end{equation}
Furthermore, $ \gamma_{1} ={\bf 1}$, so that
$ {\rm Tr} (\gamma_{1}) =n_9$ is the number
of D9-branes.

If $\gamma_{\Omega}$ is symmetric, then 
by  a unitary change of basis $\g_{\Omega} \rightarrow U \g_{\Omega} U^T$,
we can make
$\gamma_{\Omega} = {\bf 1}$. 
If $\gamma_{\Omega}$ is antisymmetric, 
 $n_9$ must be even and we can choose
a basis such that
$\gamma_{\Omega}$ is the symplectic matrix
\begin{equation}
J = \left[ \begin{array}{cc} 0&iI\\-iI&0 \end{array} \right].
\end{equation}
Lets look at the open-string sector. Tachyon is projected out by the GSO
projection. For the massless open string vector 
\bea
\psi^i_{-1/2} |0 \ ij\rangle \lambda_{ij},
\eea
the $\Omega$ eigenvalue of the
oscillator state is $-1$. \footnote{The overall sign of $ \O$ 
is fixed by the requirement that string interactions, or equivalently,
the correlation functions in the conformal field theory
preserve $ \O$. The minus sign is easiest to see for the bosonic
string \mcite{PolcTasi} and in the ghost number zero picture
for the superstring \mcite{FMS}.}
For $\gamma_{\Omega}$ symmetric, the Chan-Paton
wavefunction of the vector is then antisymmetric, giving the gauge
group $SO(n_9)$.  For $\gamma_{\Omega}$ antisymmetric, the massless vectors
form the adjoint of $USp(n_9)$.  

Tadpole cancellation requires that the integrand
in Eq.~\ref{eq:tadpole} must vanish, giving us
$ n_9=32$ and  $\g^T_\Omega = \g_\Omega$.
Therefore, the Chan-Paton wave function satisfies
$ \l^T =-\l$ and the gauge group  of Type-I theory 
is $ SO(32)$.

Let us recapitulate why we obtained $ SO(32)$, \ie,
why we needed $ 32$ D-branes. To get the number
$ 32$,  the Klein bottle diagram should be $ 32^2$ times larger
than the cylinder diagram. This factor comes  as follows. 
There is a factor of $ 2^5$ that comes
from momentum integration 
in the loop channel because of the difference between the Hamiltonians
for closed strings
and for open strings. There is an additional factor of $ 2^5$ in going
from the  variable $t$ to variable $l$ in Eq.~\ref{sums} using Eq.~\ref{treeloop}.
We see therefore that the gauge group is closely linked with the 
number of spacetime dimensions.

\section{Some Compactifications on K3}

\subsection{K3 as an Orbifold}

``Kummer's third surface''  or K3 has played an important
role  in many developments concerning duality.
Let us recall some of its properties.  $ K3$ is  a four dimensional
 manifold which has
$SU(2)$ holonomy. To understand what this means, consider a generic 4d
real manifold. If you take a vector in the tangent space at point $P$,
parallel transport it, and come back to point $P$, then, in general, it
will be rotated by an $SO(4)$ matrix:
\bea
V_i (P) \rightarrow O_{i j }\ V_i (P) \ \ \ \ O_{ij} \in SO(4).
\eea 
Such a manifold  is then said to
to have  $SO(4)$ holonomy.
In the case of K3, the holonomy is a subgroup of $SO(4)$,
namely $SU(2)$. The smaller the holonomy group, the more
``symmetric'' the space. For example, for a torus, the holonomy group
consists of  just the identity because the space is flat and 
Riemann curvature is zero;
so, upon parallel transport along a closed loop, a vector  
comes back to itself.
For a K3, there $\it{is}$ nonzero curvature 
but it is not completely arbitrary: the Riemann
tensor is non-vanishing but the Ricci
tensor $R_{ij}$ vanishes. Therefore, K3 can alternatively
be defined as the manifold of compactification that solves
the vacuum Einstein equations.

Only other thing about K3 that we need to know 
is the topological information.  A surface can have nontrivial cycles
which cannot be shrunk to a point. For example, a torus
has two nontrivial 1-cycles. 
The number of nontrivial k-cycles
which cannot be smoothly deformed into each other  is given by
the $ k$-th Betti number $ b_k$ of the surface.
The number of non-trivial $ k$-cycles is in one to one correspondence
with the number of harmonic $ k$-forms on the surface
given by the $ k$-th
de-Rham cohomology \mcite{GSW}.
A harmonic k-form $ F_k$ satisfies the Laplace equation, or equivalently
satisfies the equations
\bea
d^\ast F_k=0,\qquad d F_k =0
\eea
A manifold always has a harmonic 0-form, {\it viz.}, a constant, and a
harmonic 4-form, {\it viz.},  the volume from, assuming we can integrate on it.
K3 has no harmonic 1-forms or  3-forms, but has
 22 harmonic 2-forms. 
So, the Betti numbers for K3 are:
\bea
b_0 =1, \qquad b_1 =0,\qquad b_2 =22,\qquad b_3=0,\qquad b_4=1.
\eea
Out of the $ 22$ 2-forms, $19$ are anti-self-dual,
and $ 3$ are self-dual. In other words,
\bea
b_2^s =3,\qquad b_2^a=19.
\eea
This is all the information one
needs to compute the massless spectrum of compactifications
on K3.

K3 has a simple description as a $ \bZ_2$ orbifold of 
a 4-torus. Let $(x_1, x_2, x_3, x_4)$ 
be the real coordinates of the torus  $ \bT^4$. Let us further
take the torus to be a product
$ \bT^4 = \bT^2\times \bT^2$. Let us introduce complex coordinates
$(z_1, z_2)$, $z_1 = x_1 + ix_2$ and  $z_2 = x_3
+ ix_4$. The 2-torus with coordinate $z_1$ is defined
by the identifications $z_1 \sim z_1 +1 \sim z_1 +i$, and similarly
for the other torus.
The tangent space group is 
$Spin(4) \equiv SU(2)_1 \times SU(2)_2,$
and the vector representation is ${ \bf  4v} \equiv ( {\bf 2, 2} )$.
If we take a subgroup
$
SU(2)_1 \times U(1)
$
of $ Spin(4)$, then the vector decomposes as
\begin{eqnarray}
{\bf 4v} = {\bf 2_+} \oplus {\bf \bar 2_-}.
\eea
The coordinates
$\pmatrix{z_1,  z_2}$ transform 
as the doublet  ${\bf 2_+}$
and $\pmatrix{\bar z_1, \bar z_2}$ as
the  $\bf \bar 2_- $. 
The $ \bZ_2 =\{ 1, I\}$ is generated by
\bea
I : (z_1, z_2) \rightarrow (-z_1, -z_2).
\eea
This $\bZ_2$ 
is a subgroup and in fact the center of $SU(2)_1$.
Consequently, as we shall see, the resulting manifold has
$ SU(2)$, indeed a $ \bZ_2$ holonomy. For a torus coordinatized
by $ z_1$,
there are 4 fixed points of $z_1 \rightarrow -z_1$
(Figure~\ref{square}). 
Altogether, on $ \bT^4/\bZ_2$,
there are 16 fixed points. 
\begin{figure}[htb]
\begin{center}
\leavevmode
\hbox{%
\epsfxsize=1.0in
\epsffile{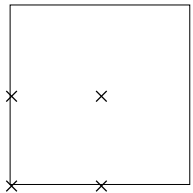}}
\caption{The four fixed point of reflection of a torus}
\label{square}
\end{center}
\end{figure}
Let us calculate the number of harmonic forms on
this orbifold. To begin with, we have on the
torus $ \bT^4$, the following harmonic forms:
\begin{eqnarray}
1&& 1\nonumber\\
4 && dx^i \ \ \ \nonumber \\
6 && dx^i \wedge dx^j \nonumber \\
4 && dx^i \wedge dx^j \wedge dx^l\nonumber \\
1 && dx^i \wedge dx^j \wedge dx^k \wedge dx^l.
\end{eqnarray}
The first column gives the number of forms indicated in the second column
where the indices
 $ i, j, k, l$ take values $ 1,\cdots 4$. 
Under the reflection $ I$, only the even forms 
$ 1, dx^i\wedge dx^j$, and $dx^i \wedge dx^j \wedge dx^k \wedge dx^l$ survive.
\bea
\begin{tabular}{llllll}
0-form & 1 & & 1 && \\
1 & 4 && 0 && \\
2 & 6 & $ { {1+ I \over 2} \atop {\longrightarrow}}$ &6 & & \\
3 & 4 && 0 && \\
4 & 1 && 1 && \\
\end{tabular},
\eea
where the  second column give the number of forms on the torus
and the third column the number of forms that survive the projection.
Let  us look at the 2-forms  from the torus that survive the $ \bZ_2$ projection.
By taking the combinations
$$
dx^i \wedge dx^j  \pm {1 \over 2} \epsilon^{ijkl} dx^k \wedge dx^l
$$
we see that three of these 2-forms are self-dual and the remaining three
are anti-self-dual.

At the fixed point of the orbifold symmetry there is a curvature singularity.
The singularity can be repaired as follows. 
We cut out a ball of radius $ R $ around each point, which has a boundary $S^3/{\bZ_2}$,
 replace it with a noncompact smooth manifold that is also Ricci flat
and has a boundary $S^3/{\bZ_2}$,
and then take the limit $ R\rightarrow 0$.
The required noncompact  Ricci-flat manifold  with
boundary $S^3/{\bZ_2}$ is known to exist and
is called the Eguchi-Hanson space \mcite{EGH}. The Betti number
of the Eguchi Hanson space are
$b_0=b_4=1$ ad $ b_2^a =1$.
Therefore, each fixed point contributes an anti-self-dual 2-form
which corresponds to a nontrivial 2-cycle in the Eguchi-Hanson
space that would be  stuck at the fixed point in the limit $ R\rightarrow 0$.

Altogether, we get $b_0=1, b_2^s =3$, $ b_2^a=3 + 16=19, b_4=1$,
and $ b_1=b_3=0$  giving us the cohomology of K3. It obviously has
$SU(2)$ holonomy. Away from the fixed point, a parallel transported
vector goes back to itself, because all the curvature is concentrated
at the fixed points. As we go around the fixed point a vector is returned
to its reflected image (for instance, $(dz_1, dz_2) \rightarrow -(dz_1, dz_2)$),
\ie, transformed by an element of $ SU(2)$.

In string theory there is no need to repair the singularity by hand.
We shall see in $ \S{5.3}$ and $\S{5.4}$ that the twisted states
in the spectrum of Type-II string moving on an orbifold
automatically take care of the repairing.
The twisted states  somehow know
about  the Eguchi-Hanson manifold that would be necessary
to geometrically repair  the singularity. 
A general method of computing the cohomology of orbifolds in
conformal field theory is described in \mcite{Zasl}.

\subsection{K3 as an Elliptic Fibration over a 2-sphere}

Let us first recall the description of a 2-torus as an elliptic curve.
An elliptic curve is a  complex 1-dimensional curve defined
by a polynomial equation involving  two complex variables $ x$ and $ y$,
\bea
\label{curve}
y^2 = x^3 + f x + g,
\eea
where $f$ and $g$ are complex numbers that determine 
the parameters of the torus.

There are a number of ways to see that this equation
defines a torus.
Before proceeding it will be useful to recall
 some relevant facts from algebraic geometry
\mcite{GSW}.
A Calabi-Yau $ n$-fold $ ({\bf CY}_n)$ is an
$ n$ complex 
dimensional
manifold with  $SU(n)$ holonomy. For example, K3 is a Calabi-Yau 2-fold,
since it has $ SU(2)$ holonomy.  
A simple way to obtain a Calabi-Yau $ n$-fold is to 
define  a special hypersurface in a weighted projective variety
of dimension $ n+1$.
Recall that a weighted projective variety of dimension $ n+1$,
is defined by 
$ n+2$
complex coordinates
$(z_1, z_2, \ldots, z_{n+2})$, not all zero,
subject to the an equivalence relation
\bea
(z_1, z_2, \ldots, z_{n+2}) \sim (\l^{r_1}z_1, \l^{r_2}a_2, 
\ldots, \l^{r_{n+2}}z_{n+2})
\eea
where $ \l$ is any nonzero complex number, and
the integers $ r_1, \ldots, r_{n+2}$ are called the weights.
The  projective variety   defined in this manner is denoted by
${\bf WP}^{n+1}_{r_1, \ldots, r_{n+2}}(z_1, \ldots, z_{n+2})$.
For example, consider $ {\bf WP}^1_{1, 1}(z_1, z_2)$, 
also known as $ \bC\bP^1$. It is defined
by two complex coordinates satisfying the equivalence
relation $ (z_1, z_2) \sim \lambda (z_1, z_2)$.
When $ z_2\neq 0$, we  choose $ \lambda = 1/z_2$ so
that $(z_1, z_2)$ is equivalent to $ (w_1, 1)$ with
 $ w_1 =z_1/z_2$. The points $ (w_1, 1)$ define the complex
plane. When $ z_2=0$, we choose $ \l =1/z_1$ 
so that $(z_1, 0)$ is equivalent to $ (1, 0)$. This additional
point $ (1, 0)$ can be regarded as the `point at infinity' that needs
to be added to `compactify' the  $ w_1$ complex plane to get the Riemann 
sphere.
In general $ {\bf WP}^{n+1}$ gives a suitable compact $ n+1$-dimensional
complex manifold. A hypersurface
defined by the vanishing of a complex homogeneous
polynomial equation of degree $ k$
would be a  complex submanifold of dimension $ n$.
To obtain  a Calabi Yau manifold there is an additional
requirement: the degree of the polynomial
 $k$ must equal the sum of the weights of the coordinates.

In summary, a Calabi-Yau manifold is defined by a homogeneous
polynomial of degree $ k$,  in a weighted projective variety,
${\bf WP}^{n+1}_{r_1, \ldots, r_{n+2}}(z_1, \ldots, z_{n+2})$ such that
\bea
\label{chern}
k =\sum_{i=1}^{n+2} r_i.
\eea
If  Eq.~\ref{chern} is satisfied, then the first Chern class of the
hypersurface vanishes ensuring that it has $ SU(n)$ holonomy.

Let us return to the torus after this digression.
A torus is the simplest Calabi-Yau manifold, \viz, ${\bf CY}_1$. 
It has $ SU(1)$ holonomy,
that is to say, no holonomy at all. In other words it is flat.
In particular we see that the equation
\bea
w y^2 = x^3 + f w^2 x +  g w^3 
\eea
defines it as a cubic in 
$ {\bf WP}^3_{1, 1, 1}(w, x, y)$
which obviously satisfies
 Eq.~\ref{chern}.
For $ w\neq 0$ we can scale it out to get
Eq.~\ref{curve}.
The point $w=0$ is the point at infinity which is needed to ensure compactness.

A more geometric way to recognize Eq.~\ref{curve} as a torus
is to note that 
\bea
y = \pm \sqrt{(x-c_1)(x-c_2)(x-c_3)}
\eea
is a function of $x$ defined on the double cover the Riemann sphere
with four branch points: $c_1, c_2, c_3$ and the point at infinity.
{}From Figure~\ref{branch} we recognize that the resulting Riemann surface
has  the topology
of a torus.
\begin{figure}[htb]
\begin{center}
\leavevmode
\hbox{%
\epsfxsize=4.5in
\epsffile{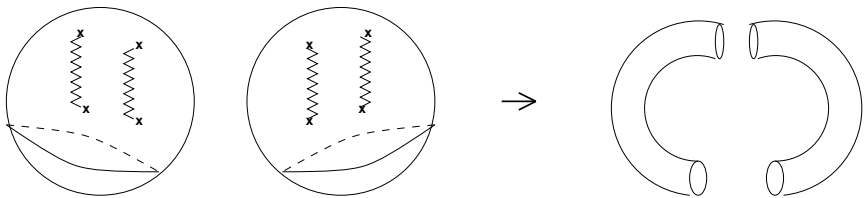}}
\caption{A double cover of branched sphere.}
\label{branch}
\end{center}
\end{figure}
In equation~\ref{curve} the parameters $ f$ and $ g$ are constants.
The modular parameter $ \tau$ of the torus which determines
the shape of the torus up to conformal rescaling (Fig~\ref{torus})
is given by the elliptic $ j$ function.
\begin{figure}[htb]
\begin{center}
\leavevmode
\hbox{%
\epsfxsize=0.8in
\epsffile{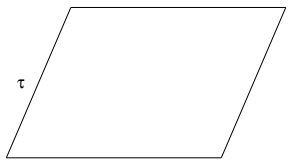}}
\caption{A torus with modular parameter $ \tau$}
\label{torus}
\end{center}
\end{figure}
The modular parameter
$\tau$ of the torus is given by
\bea
j(\tau) = {4 (24 f)^3 \over 27 g^2 + 4f^3}
\eea
where $j(\tau)$ is the well known j-function, 
\bea
\label{j}
j(\tau) = {\theta^8_1 (\tau) + \theta^8_2 (\tau) + \theta^8_3
(\tau)^3 \over \eta (\tau)^{24}},
\eea
where
$\theta_i, i=1, 2, 3$,  are the well-known Jacobi $\theta$ functions
\mcite{Erde}
and  $\eta$ is the
Dedekind $\eta$ function
\bea
\eta(\tau) = q^{1/24} \prod_n (1-q^n), \qquad q=\exp{2\pi i\tau}.
\eea
The $j $ function gives a one-one map from 
 the fundamental domain (the moduli space of the torus) to
the complex plane. Given $ f$ and $ g$,
we can obtain
$ \tau$ by inverting $ j$.

To obtain a 2 complex dimensional fiber bundle with
an elliptic curve as a fiber and  a sphere
as a base,  locally
we can take the parameters $ f$ and $ g$ of the elliptic curve
in Eq.~\ref{curve} to be functions
of a complex coordinate $ z$ that takes values in $\bC\bP^1$.
Then for every value of $z$ i.e.
at each point, we get a torus. As we move around on the base by varying
 $z$, the parameters $f$ and $g$ 
and consequently the modular parameter $\tau$ of
the fiber will vary. The functions $f$ and $g$ should
be so chosen that  globally we 
obtain a smooth K3 that is
elliptically fibered.  
This is  achieved by taking $f(z)$ to be an arbitrary polynomial in $z$
of degree 8 and $g(z)$ to be an arbitrary 
 polynomial  of degree 12.  Given a polynomial
$f(z) = \sum_{k=0}^8 \a_k z^k $ we can define 
$f(w, z) = \sum_{k=0}^8 \a_k w^{8-k} z^k$ which is
a polynomial of degree 8 that is homogeneous
if we assign weight $1$ to both $w$ and $z$. 
The coordinates $w$ and $z$ are nothing but the projective
coordinates of $\bC\bP^1$
Now, the equation
\bea
\label{k3}
y^2 = x^3 + f (z, w) x + g (z, w)
\eea
 is a polynomial of degree 12 in 
\bea
{\bf WP}^3_{1, 4, 6, 1} (w, x, y, z) 
\eea
The sum of weights $1+ 4+6+1=12$ equals the degree
of the polynomial showing that the first Chern class vanishes. 
Hence Eq.~\ref{k3}
defines a $\bf CY_2$, \ie, a K3.
Using the equivalence relation 
$
(w, x, y, z)  \sim (\lambda w, \lambda^4 x, \lambda^6 y, , \lambda z)
$
 we can set $w=1$ 
by choosing $\lambda = (1/w)$ when $w \neq 0$
to get
\bea
\label{elliptick3}
y^2 = x^3 + f(z) x + g (z).
\eea
We shall use Eq.~\ref{elliptick3} as the defining
equation of a K3 elliptically fibered over
$\bC\bP^1$. 
\subsection{Type IIB string on K3}

Consider II-B
compactified on K3. The resulting  theory in the remaining
6-dimensional Minkowski space has  $(0, 2)$ chiral supersymmetry.
To discuss the spectrum
let us recall that  massless states are labeled by the representations
of the little group in six dimensions which is $Spin(4)=SU(2) \times SU(2)$.
With  $ (0, 2)$ supersymmetry, only two massless supermultiplets 
are possible.
In terms of representations of the little group the supermultiplets 
are given by

1. The gravity multiplet:

\qquad a graviton $({\bf 3}, {\bf 3})$,

\qquad five self-dual 2-forms $5({\bf 1}, {\bf 3})$,

\qquad gravitini $ 4({\bf 2}, {\bf 3})$,

2. The tensor multiplet:

\qquad an anti-self-dual 2-form $({\bf 3}, {\bf 1})$, 

\qquad fermions $4({\bf 2}, {\bf 1})$, five scalars $({\bf 1}, {\bf 1})$.\\
The gravitini  are right-handed whereas the fermions
in the tensor multiplets are left-handed. 

We can explicitly work out the spectrum of Type-IIB
on a K3 that is a $ \bZ_2$ orbifold. Let us take $X^m, m=6, 7, 8, 9$
to be the coordinates of the internal torus and  $X^i, i= 2, 3, 4, 5$,
to be the noncompact light-cone coordinates.
It is convenient to decompose the little group in ten dimensions
$ SO(8)$ as
\bea
Spin(8) &\supset& Spin(4)_I \times Spin(4)_E \nonumber\\
&\equiv&
SU(2)_{1I} \times SU(2)_{2I}\times SU(2)_{1E} \times SU(2)_{2E},
\eea
where the subscript $ I$ is for internal,  $E$ is for external.
With this embedding, the representations decompose as
\bea
{\bf 8v}\ =\,\ ({\bf 4v}, {\bf 1}) \oplus ({\bf 1}, {\bf 4v})  
\ \, &\equiv& ({\bf 2}, {\bf 2}, {\bf 1}, {\bf 1}) \oplus ({\bf 1}, {\bf 1}, {\bf 2}, {\bf 2}),\nonumber\\
{\bf 8s}\ =\ ({\bf 2s}, {\bf 2s}) \oplus ({\bf 2c}, {\bf 2c}) 
&\equiv& ({\bf 2},  {\bf 1}, {\bf 2}, {\bf 1}) \oplus ({\bf 1}, {\bf 2}, {\bf 1}, {\bf 2}),\nonumber\\
{\bf 8c}\ =\ ({\bf 2s}, {\bf 2c})\oplus ( {\bf 2c} , {\bf 2s})  
&\equiv& ({\bf 2}, {\bf 1}, {\bf 1}, {\bf 2}) \oplus ({\bf 1}, {\bf 2}, {\bf 2} , {\bf 1}).
\eea
The orbifold group is a $ \bZ_2$ subgroup of $SU(2)_{LI}$
which acts as $-\bf 1$ on the doublet representation $ \bf 2$.

Untwisted sector:\\
The states in the untwisted sector are obtained by
keeping  the $ \bZ_2$ invariant states
of the original 10-dimensional states.
\bea
({\bf 8v}\oplus {\bf 8c}) \otimes ({\bf 8v}\oplus {\bf 8c}).
\eea
For example, the bosons (labeled by $ SU(2)_{1E} \times SU(2)_{2E}$
quantum numbers are
\bea
&& [4({\bf 1}, {\bf 1})\otimes 4( {\bf 1}, {\bf 1})] 
\oplus [({\bf 2}, {\bf 2})\otimes ( {\bf 2}, {\bf 2})]\nonumber\\
&& 
[2({\bf 1}, {\bf 2})\otimes 2({\bf 1}, {\bf 2}) ] 
\oplus [2({\bf 2}, {\bf 1})\otimes 2({\bf 2}, {\bf 1}) ]
\eea
This gives rise to a graviton, $25$ scalars, $5$ self-dual and 5 anti-self-dual
2-forms. The fermions can be obtained similarly which give the superpartners
required by supersymmetry. Together, we get the gravity multiplet
and five tensor multiplets.

Twisted Sector:\\
There are 16 twisted sectors coming from the 16 fixed points.
The bosonic fields and fermionic fields are twisted according
to their transformation property under the $\bZ_2$. We see 
from that four fermions that transform as $ 2({\bf 2} , {\bf 1})$
and four bosons that transform as $ ({\bf 2}, {\bf 2})$ are $ \bZ_2$
invariant and are not twisted where as the four other are twisted.
The ground state energy is zero because there are equal number
of bosons and fermions that are twisted. The untwisted fermions
have zero modes. By steps analogous to those
that led to Eq.~\ref{rep} in $ \S{2.1}$,
the zero mode algebra gives rise to
a four dimensional representation 
$ ({\bf 2} , {\bf 1}) \oplus 2({\bf 1} , {\bf 1})$.
Therefore the massless representation is
\bea
[({\bf 2} , {\bf 1}) \oplus 2({\bf 1} , {\bf 1}) ]\otimes
[({\bf 2} , {\bf 1}) \oplus 2({\bf 1} , {\bf 1})]
\eea
which gives precisely the particle content of a tensor multiplet.
Therefore, the twisted sector contributes 16 tensor multiplets.

The massless spectrum of Type-IIB on a K3 orbifold
thus consists of  a gravity multiplet and  21 tensor multiplet
together from the untwisted and the untwisted sector.
There are $105$ scalars that parametrizes the moduli space
$O( 21, 5; \bZ)\backslash O(21, 5; \bR)/O(21; \bR)\times O(5; \bR)$.

The spectrum of Type-IIB is chiral. 
A chiral theory can have gravitational anomalies
In $ 4k+2$ dimensions 
Up to overall normalization
the gravitational anomalies are
\bea
\label{anomalies}
\label{anomaly}
I_{3/2} &= -\frac{43}{288}(tr R^2)^2 + \frac{245}{360} tr R^4, \cr
I_{1/2} &= +\frac{1}{288}(tr R^2)^2 + \frac{1}{360} tr R^4, \cr
I_{A} &= -\frac{8}{288}(tr R^2)^2 + \frac{28}{360} tr R^4.
\eea
Here $I_{3/2}$, $I_{1/2}$, and $I_A$ refer to the anomalies for
the gravitino, a right-handed fermion,
and a  self-dual two-form $(1, 3)$ respectively
\footnote{In $ 4k+2$ dimensions, the CPT
conjugate of a left-handed fermion is also left-handed.
Therefore,  gravitational couplings can be chiral
and consequently  gravitational anomalies are possible.
Contrast this with the $ 4k$ dimensions as in the 
 familiar case of four dimensions where the CPT conjugate
of a left-handed fermion is right-handed and a CPT-invariant
theory is automatically nonchiral unless there are gauge charges in addition
to gravity that distinguish between left and right.}.

To get a consistent theory, the total gravitational anomalies
must cancel.
This requirement is very restrictive and in fact  
completely determines the spectrum for the theory with
(0, 2) supersymmetry. 
Using  the formulae from Eq. \ref{anomalies} it is easy
to check that gravitational anomalies cancel only when there are precisely
$ 21$ tensor multiplets along with the gravity multiplet \mcite{ToSe}. 
This, as we have seen, is the spectrum of II-B compactified on K3.

\subsection{Type IIA string on K3}

Type-IIA compactified on a K3 gives a non-chiral theory
in six-dimensional  Minkowski spacetime with $ (1, 1)$ supersymmetry.
There are only two supermultiplets that are possible.

1. The gravity multiplet:

\qquad a graviton $({\bf 3}, {\bf 3})$, a scalar $({\bf 1}, {\bf 1}) $,

\qquad four vectors $4({\bf 2}, {\bf 2})$, 
a 2-form $({\bf 3}, {\bf 1}) \oplus ({\bf 1}, {\bf 3})$,

\qquad gravitini $ 2({\bf 2}, {\bf 3})\oplus 2({\bf 3}, {\bf 2})$

\qquad two fermions $2({\bf 2}, {\bf 1}), 2({\bf 1}, {\bf 2})$

2. The vector multiplet:

\qquad a vector $({\bf 2}, {\bf 2})$,

\qquad four scalars $4({\bf 1}, {\bf 1})$,

\qquad gauginoes $2({\bf 2}, {\bf 1}), 2({\bf 1}, {\bf 2})$. 

The spectrum can be found as in the previous section.

Untwisted sector:\\
Now the ten-dimensional states are 
\bea
({\bf 8v}\oplus {\bf 8s}) \otimes ({\bf 8v}\oplus {\bf 8c}).
\eea
Keeping   $ \bZ_2$ invariant states
we obtain the gravity multiplet and 4 vector multiplet.

Twisted Sector:\\
Now, the fermions that not twisted have different quantum number
on the left and on the right. Therefore, the representation 
of the fermion zero mode algebra is different on the left and the right.
The massless representation are given by the product
\bea
 [({\bf 1} , {\bf 2}) \oplus 2({\bf 1} , {\bf 1})]\otimes
 [({\bf 2} , {\bf 1}) \oplus 2({\bf 1} , {\bf 1})]
\eea
which gives precisely the particle content of a vector multiplet.
Therefore, the twisted sector contributes 16 tensor multiplets
one from each of the fixed points.

The massless spectrum of Type-IIA on a K3 orbifold
thus consists of  the gravity multiplet and  20 vector multiplets.
There are $ 80$ scalars that parametrize the moduli space
$O( 20, 4; \bZ) \backslash O(20, 4; \bR)/O(20; \bR)
\times O(4; \bR)$.

\subsection{ F-theory on K3}

Until recently string compactifications basically solved vacuum Einstein
equations in the low-energy limit for some compact manifold K,
\bea
R_{ij} = 0, 
\eea
In fact, unbroken supersymmetry in the
remaining noncompact dimensions requires that the compact manifold
be a Calabi-Yau manifold with   $SU(n)$ holonomy, in particular
with vanishing first Chern class.

Instead of solving the vacuum Einstein equations, one can imagine
solving the equations with some nonzero background fields.
In particular, one can ask if there are consistent solutions
of the Type-IIB string where the 
 complex  field $\l = \chi + ie^{-\phi}$ varies.
When all other massless fields of Type-IIB theory are set to zero,
the equations of motion for the graviton $ g_{MN}$ and the scalar $ \l$ can
be derived from the action
\bea
\label{actionthree}
\int  d^{10} x \sqrt{g} \left(R - {1 \over 2} {g^{MN }\partial_M \bar\l
\partial_N\l \over Im (\tau)^2}\right).
\eea

A particularly interesting nontrivial solution of this action is Type-IIB
compactified on a 2-sphere ($S^2\equiv  \bf{CP}^1$). The  spacetime of this
compactification is
of the form 
\bea
\bM^8 \times S^2,
\eea
where $ \bM^8$ is flat Minkowski spacetime with coordinates
$ X^0, X^1,\ldots, X^7$ and $ S^2$ is the compactification
sphere with coordinates $ X^8, X^9$.
Now,  the sphere which has nonzero curvature
$ R_{ij}\neq 0$. In fact, the first Chern class of $S^2$ is the Euler
character of the sphere which is nonzero, so $S^2$ is obviously
not a Calabi-Yau manifold. The way equations of motion are still satisfied
is that the spacetime
 contains 24 7-branes. A 7-brane, as we shall see,
can be thought of a special topological defect which couples to $ \lambda$.
The worldvolume of the 7-brane fills the noncompact $ \bM^8$,
so in the transvers $ S^2$ it looks like a point.
The energy momentum tensor $ T_{ij}^\l $ of $ \l$ and the metric
are  precisely such that they
solve  the Einstein equation
\bea
R_{ij} -\half g_{ij} R= T_{ij}^\l.
\eea

To describe the solution, let
 us first discuss a single 7-brane. Let $z=X^8 +iX^9$ be the complex
coordinate on the plane transverse to the 7-brane. 
The coordinates $ X^0, X^1,\ldots, X^7$
are along the worldvolume of the 7-brane. The equation
of motion for $ \l$ that follows from Eq.~\ref{actionthree}
is
\bea
 (\l -\bar\l )\partial\bar\partial \l -2\partial \l\bar\partial \l =0.
\eea
This equation is solved by any $ \l$ that is a holomorphic
function of $ z$
\bea
\bar \partial \l( z, \bar z) =0.
\eea
Not any holomorphic function will do. Recall that $ \l$ parametrizes,
after $ SL(2, \bZ)$ identification, the fundamental domain of the
moduli space of a torus (Figure~\ref{fig:modular}).
To get a well-defined solution we  want a one-to-one map from the fundamental
domain to the complex plane. We have already seen
that the $ j$-function defined in Eq.~\ref{j} gives precisely such a map.
Therefore, instead of looking for $ \l$ as a function of $ z$ it is
convenient to look for $ j(l)$ as function of $ z$.
Furthermore, the resulting
configuration should have finite energy to be an acceptable
solution.

The simplest solution that satisfies all the requirement
is 
\bea
j(\l(z)) = {1\over z}
\eea
which has the right properties.

If we suppress $ 6 $ of the coordinates along the 7-brane
(say $ X^2, \ldots, X^7)$ then the 7-brane looks like a cosmic
string in four dimensions $ X^0, X^1, X^8, X^9$. This 
in fact is nothing but the ``stringy'' cosmic string solution
discussed by Greene {\it et. al.}\mcite{GSVY}.
Near $ z=0$ $ j$ has a pole. The only pole of $ j$
is at $ q=\exp{2\pi i\l}=0$ at $ \l_2\rightarrow \infty$.
For large $ \l_2$  we have,
\bea
j(\l) \sim \exp{-2\pi i\l}.
\eea
The solution looks like
\bea
\l ={1\over 2\pi i}\log{z}
\eea
near $ z=0$.

If we go around the origin on a circle at infinity in the $ z$ plane
 with  $ z\rightarrow  z e^{2\pi i}$ then $ \l\rightarrow  \l+1$.
This is very much like a global cosmic string or a vortex line
in superfluid helium. A cosmic string is a
topological defect in which the phase angle $\theta$ of the
order-parameter field has a winding number. 
\begin{eqnarray}
\theta (ze^{2\pi i}) \rightarrow& \theta (z) + 2 \pi
\eea
The RR-field $ a$ in Type-II string is very similar
to the phase of the order parameter $\theta/2\pi$.
An important difference is that the total energy of 
global string or a vortex line  has an infrared
divergence because very far from the core the superfluid in
a vortex undergoes huge rotation. By contrast, the
energy density of the stringy cosmic string is finite.
This is possible because the $ \l$ field can undergo
$ SL(2,\bZ)$ jumps away from the core.
Near the core $ z=0$, the $ \l$ field has a nontrivial
monodromy or jump under the element $ T$ of $ SL(2, \bZ)$,
$ T: \l \rightarrow \l +1$,
but far away it can undergo jumps under other elements 
of $ SL(2, \bZ)$.

The nontrivial monodromy of $\l$ around
the point $ z=0$ means in string theory that there is a 7-brane at this point
that is magnetically charged with respect to the scalar $\l$. 
Indeed, near the origin, this is exactly like a  D7-brane in Type-IIB
which is magnetically charged with respect to the RR scalar.
In other words, it couples to the 8-form RR potential $A_8$ that is dual to 
$a$,
\bea
dA_8 = \ast da.
\eea

Let us now look at the effect of the 7-brane on the metric. 
Because of the energy density contained in the field $ \l$,
the metric in $ z$ plane has a conical deficit near $ z=0$
with conical deficit angle $ \d$.
The metric near such a point $z_i$ can be
found explicitly \mcite{GSVY}, and has the form
\bea
\label{metric}
ds^2 = {dz d\bar z \over |z - z_i|^{1/6}} \sim r^{-1/6} (dr^2 + r^2
d\psi^2), \qquad 0 \leq \psi \leq 2\pi.
\eea
A metric of the form
\bea
r^{-2\lambda} (dr^2 + r^2 d\psi^2)
\eea
can be written in the form
\bea
 \ dt^2 + t^2 (1 - \lambda)^2 d\psi^2,  
\eea
with $ t={r^{1-\lambda} / (1 - \lambda)}$.
We can nowredefine the angle
\bea
\phi = (1-\lambda) \psi
\eea
to bring it to the standard flat metric of the plane in polar coordinates
$ dt^2 + t^2 d\phi^2$. But then $ \phi$
goes from $0$ to $2\pi - 2\pi \lambda$.
Therefore, the deficit angle is 
$ \d= 2\pi \lambda $ which for the metric in Eq.~\ref{metric}
is $  \pi/6$.
The deficit angle measures the total curvature 
or $\d = \int R$ where $ R$ is the Ricci scalar. 

So far, $ z$ is a coordinate on  the noncompact complex plane.
If we put precisely 24 7-branes on the plane then 
the plane
 curls up
into a sphere.
This is because the total contribution to the deficit angle from all 
the cosmic strings now
adds up to  $ 4\pi$ to make up the solid angle of a sphere.
For the sphere, the Euler
character $ 2$ and  $ \int R = 2\pi \chi =4\pi$.

A sphere with a collection of $ 24$ F-theory 7-branes is a compact
manifold. In fact, one can associate with it
 an elliptically fibered K3, with the sphere 
as the base, if the $ \tau$ parameter of the fiber torus
is identified with $ \l$. 
The elliptic $ K3$ is described
by Eq.~\ref{elliptick3} in $ \S{5.3}$
\bea
y^2 = x^3 + f(z) x + g (z).
\eea
where $ f$ and $ g$ are polynomials  of degree $ 8$ and $ 12$ respectively.
The locations of  7-branes are  determined by 
the zeroes  of  the discriminant $\Delta \equiv 4f^3 + 27g^2$
which is the denominator of the $ j$ function.
Since $ f$ is a polynomial of order $ 8$ and $ g$ is a polynomial
of order $ 12$,
there will, in general,  be {24} zeroes of the discriminant 
which correspond to the locations of the 24
7-branes.
Near every zero $ z_i$, we have
\bea
j(\l) \sim {c \over (z-z_i)} \sim {1 \over q}
\eea
\bea
\l (z) \simeq {1 \over 2\pi i} \log (z-z_i),
\eea
corresponding to a single 7-brane.

The compactification of Type-IIB of $S^2$
that we have described has been called an `F-theory' compactification on K3. 
`F-theory' refers to a possible 12-dimensional theory which when
compactified on $ \bT^2$ would give Type-IIB.
In general,
consider an elliptically fibered Calabi-Yau manifold $K$ which is a fiber
bundle over a base manifold $B$ with a torus as a fiber whose
complex structure parameter is $\t$. Even-though $K$ is a smooth
manifold, there will be points in the base manifolds where
the fiber becomes singular, and the parameter $\t$ can have
a nontrivial $SL(2, Z)$ monodromy around these points. An
F-theory compactification on $K$ refers to a compactification
of Type-IIB theory on $B$, where the coupling $\l$ is
identified with $\t$.

\section{ Applications of Orientifolds to Duality}

\subsection{General Remarks on Duality}

Apart from T-duality,  two other dualities   will be relevant
to us in the following discussion. Both
are  `strong-weak' dualities which
relate the strong coupling limit of a ten-dimensional string theory 
to the weak coupling
limit of the dual theory.
 
(1) Duality between the Type-I string and  the $ SO(32)$ Heterotic string.\\
The massless spectrum as well as the form of the low energy
effective action agrees \mcite{WittStri} under the duality transformation which
takes the coupling constant of Type-I string to the inverse coupling constant
of the Heterotic string. The spectrum of
some of the solitons  in the two theories has  also been checked  to be in agreement with duality \mcite{Dabh,Hull,PoWi}.

(2) Self-duality of the Type-IIB string.\\
We have already discussed this. The element $ S$ of the $ SL(2, \bZ)$ duality
group takes the coupling constant to the inverse coupling constant
in the dual theory.

For a detailed discussion of the evidence for these  dualities,
see \mcite{SenNonp}.
In this section we would like to use these  dualities,
T-duality, and our knowledge of orbifolds and orientifolds,
to deduce two more dualities in lower dimensions.

\subsection{Duality of Type-IIA on $K3$ and Heterotic on $T^4$}

The main principle that we use in this subsection is 
`fiberwise application of duality', which we explain below.

Consider a theory $ A$
compactified on $K_A$ that is dual to another theory
$ B$ compactified on $K_B$. 
This duality can be used to deduce some further dualities.
Consider $ A$ and $B$ compactified 
respectively on $E_A$ and $E_B$, which are obtained by fibering $K_A$
and $K_B$ over $\Sigma$.
By this, we mean that  locally $E_A$ looks like 
$K_A \times \Sigma$. The moduli $ m_\a$ of the fiber $K_A$
can vary as a function of the coordinates of the base manifold
$ \Sigma$. 
As long as the moduli of $K_A$ vary slowly we expect
to be able to use the original duality to derive a new duality
between $ A$ on  $E_A$ and $ B$ on $E_B$. 
There are two possibilities that aries.

(i) The first possibility is that the fiber is  smooth at all points on 
the base manifold $ \Sigma$. In this case, the
duality
between $ A$ on  $E_A$ and  $ B$ on $E_B$ follows from the
`adiabatic argument' of Vafa and Witten \mcite{VaWi}. The idea is that 
we can choose the size of $ \Sigma$
to be very large. Then locally $ A$ compactified on $ E_A$
has $ K_A\times \bM^n$ as the target spacetime and 
and $ B$ on $ E_B$ has $K_B\times \bM^n$ as the target spacetime.
Knowing the duality  between the two,
we can assert the new duality.
The fibered structure will become apparent to a local observer
only after circumnavigating
the (very large) manifold $\Sigma$ and so will not be relevant
to local physics. We can thus establish the duality
between $ A$ on  $E_A$ and $ B$ on $E_B$
 in the limit of large $ \Sigma$.
Now, if  we adiabatically reduce the
volume of $\Sigma_2$ we expect that the duality 
will continue to hold.

(ii) The second possibility is that the fiber is smooth 
everywhere on $ \Sigma$ except at a few discrete points 
where it degenerates ( Figure~\ref{fig:degen}).
\begin{figure}[htb]
\begin{center}
\leavevmode
\hbox{%
\epsfxsize=0.7in
\epsffile{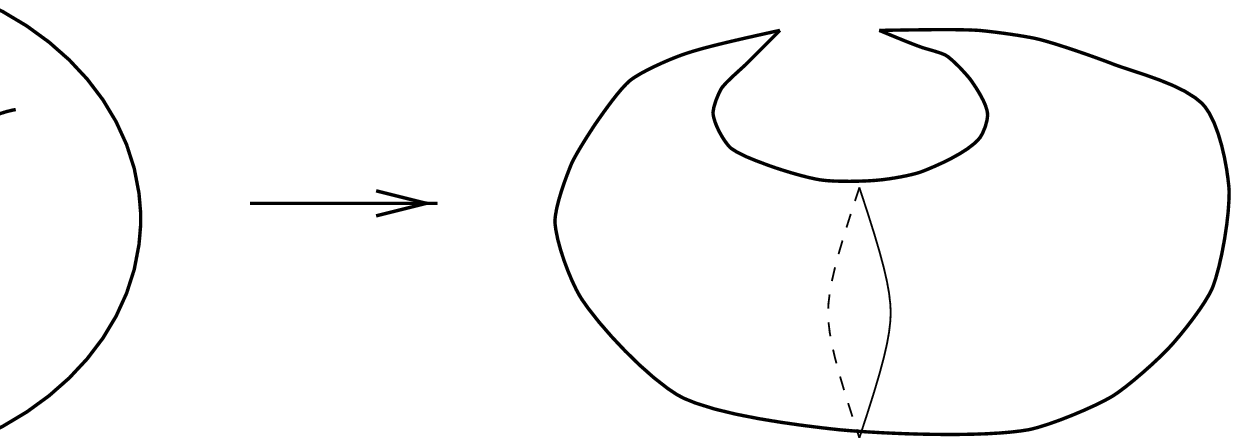}}
\caption{A torus degenerates into a sphere by pinching a handle.}
\label{fig:degen}
\end{center}
\end{figure}
The total manifold $E_A$ can still be smooth,
it is only the fibration that becomes singular.
In this case, the adiabatic argument is strictly not applicable.
Near the points where the fiber degenerates,
 the argument breaks because the moduli vary rapidly.
However,  in a large number of examples constructed so far, 
the resulting theories
do appear to be dual as long as
the number of singular points is  a set of ``measure zero''.
Heuristically,  even with the singular points, duality 
is forced by the duality in the bulk.

It is instructive to apply these arguments to the special
case of fibered manifolds that are orbifolds.
Take a smooth manifold $\CM$  with a $ \bZ_2$ symmetry $\{1, s\}$. 
Let $K_A$ and $K_B$ have some $\bZ_2$ 
symmetry $\{1, h_A\}$ and $\{1, h_B\}$. Take
\bea
E_A= {K_A \times \CM \over \{1, sh_A\}}, \qquad 
E_B = {K_B \times \CM \over \{1, sh_B\}}
\eea

Here there are two possibilities. 

(i) If $ s$ has no fixed points  on $ \CM$,
then we have the possibility (i) above.  
The orbifold $ E_A$ can
be viewed at all points as a fiber bundle with $\Sigma = \CM/\{1, s \}$ as
the base space 
and $ K_A$ as the fiber. The fiber
is smooth everywhere.
There is a twist $ h_A$ along the fiber
as we move along a closed curve from a point  $ p$ on $ \Sigma$
and its image $ s(p)$. 

(ii) If $ s$ leaves some points on $ \CM$ fixed,
then we have the possibility (ii).
The orbifold  $ E_A$ still has the structure
of a fiber bundle with  base manifold  $\Sigma = \CM/\{1, s \}$
and fiber $ K_A$ everywhere except the fixed points.
At  the fixed points the fiber degenerates to  $ K_A/\{1, h_A\}$
giving us a singular fibration.
\footnote{There is a third possibility that $ s$ leaves the entire
manifold $ \CM$ invariant. In this case the resulting orbifolds
do not have the fibered structure even locally. In such a situation
orbifolding does not commute with duality \mcite{SenDual,SenOrbi}
and cannot be used to deduce new dualities. 
For example, Type-I is an orientifold of Type-IIB by
$ \{1, \O\}$
and Type-IIA is an
orbifold of Type-IIB by $ \{ 1, \FL\}$. Now,
$ \O$ is conjugate to $ \FL$ by the element $S$ of the
$ SL(2,\bZ)$ group but Type-I is $ \it not$ dual to Type-IIA.}

As an illustration of the second possibility we now `derive' the duality
between Type-IIA on $K3$ and Heterotic on $T^4$.
Let us take  $A$ to be the Type-IIB theory in ten dimensions and $ B$ to be the
Type-IIB theory that is S-dual
to $ A$ ($ K_A$ and $ K_B$ are null sets).
Compactify both sides on a torus $ \bT^4$
with  coordinates $X_6, X_7, X_8, X_9$,
\ie, $ \CM = \bT^4$.
Now, take $ h_A =\FL$, $ h_B=\O$, and $ s$ to be the
reflection
$$
I_{6789} : (X_6, X_7, X_8, X_9) \rightarrow (-X_6, -X_7, -X_8, -X_9)
$$
We have used the duality relations
\bea
IIB \arrowtop{S}  IIB, \qquad  \FL \arrowtop{S} \O,
\eea
from $ \S{2.3}$.
Therefore, from the earlier arguments, we get the duality
\bea
\label{eq:dualityone}
{\rm IIB~~on}~~{\bT^4 \over \{1, \FL I_{6789}\}} \ \equiv \ 
{\rm IIB~~on}~~ {\bT^4 \over \{1, \O I_{6789}\}}
\eea
Now we can  use T-dualities to turn these orientifolds 
into more familiar ones. We use two observations.

(1) If we T-dualize one of the coordinates, say  $X^6$, 
we get Type-IIA theory on the dual circle. Moreover,
the symmetry  $ I_{6789}\FL$ in IIB maps under this T-duality
onto $ I_{6789}$ in the dual IIA. To see this, 
recall that if we  T-dualize in a direction $ X^i$
\bea
(X^i_L, X^i_R)\rightarrow (-X^i_L, X^i_R)
\eea
then the left-moving fermions transform as  
\bea T_i:\quad
S^a\rightarrow P_i S^a,\eea
where $P_i= \G\G^i$. 
Now using the properties of the $ \G$ matrices,
we see that $ P_i$ does not square to identity but instead
$P_i^2 =\FL$. Furthermore $P_i P_j =-P_j P_i$.
Note that the reflection $I_{6789}$ acts
as $P_6P_7P_8P_9 S^a$ on the left-moving fermions,
and similarly on the right-moving fermions. Using the properties
of $ P_i$ matrices we see that
\bea
I_{6789}\FL\ = \ {T_6}^{-1}\, I_{6789}\, T_6
\eea
Therefore, the orbifold  Type-IIB on ${\bT^4 / \{1, \FL I_{6789}\}}$
is T-dual to Type-IIA on the  K3 orbifold $ {\bT^4 / \{1,  I_{6789}\}}$.

(2) T-duality of  all four coordinates maps
$ I_{6789}\O$ in IIB
into $\O$ in IIB by a reasoning similar to the above.
\bea
I_{6789}\O= \ {T_{6789}^{-1}\, \O\, T_{6789}}
\eea
Hence, Type-IIB on the orientifold ${\bT^4 / \{1, \O I_{6789}\}}$
is 
T-dual to Type-IIB on the orientifold ${\bT^4 / \{1, \O \}}$,
which is nothing but Type-I on $ \bT^4$.
From the duality between Type-I and Heterotic in ten
dimensions, it is dual to Heterotic on $ \bT^4$.

We conclude from Eq.\ref{eq:dualityone} and points (A) and
(B) above that
\bea
\label{eq:dualitytwo}
{\rm IIA~~on}~~{\rm K3} \ \equiv \ 
{\rm Heterotic ~~on}~~ {\bT^4 }.
\eea
This equivalence  has been established at a
special point in the moduli space where
the K3 becomes the orbifold $ \bT^4/\bZ_2$.
The duality gives a one-one map between the massless fields
on both sides. By giving expectation values to the massless scalar
field we can move around in the moduli space and establish
the duality at all points in the moduli space.

Thus, starting with the  $SL(2, \bZ)$ duality of
II-B with 32-supercharges (and the duality between Type-I and
Heterotic), we can get all the structure of the more
interesting `string-string' duality \mcite{HuTo}
between II-A on K3 Heterotic on $ \bT4$ with
16-supercharges. This is quite an explicit construction, and all we
needed to do was to keep track of a few discrete symmetries
and follow the orientifold and orbifold construction.

Let us quickly check if the spectrum of Heterotic on
$\bT^4$ matches with the dual Type-IIA spectrum.
The moduli space of Heterotic on $ \bT^4$ is the
Narain moduli space \mcite{Nara}
$O( 20, 4, \bZ) \backslash O(20, 4, \bR)/O(20, \bR)
\times O(4, \bR)$
which is identical to the moduli space of Type-IIA on K3.
At a generic point in the moduli space, the gauge
group $ SO(32)$ is broken to $ U(1)^{16}$. In six
dimensions we get 4 vector bosons $ g_{\mu m}$,
4 vectors from $B_{\m m}$,  
and 16 vectors from the
original gauge fields in ten dimensions,
$ A_\mu^I$ ($I=1, \ldots, 16$ is the gauge index, 
$m=6, \ldots, 9$ is the internal index, and 
$ \m$ is the Minkowski index  $ \mu=0. \ldots, 5$).

Altogether there are $24$
vector bosons, exactly as in the case of Type-IIA ($ \S{5.4}$), 
which transform in the vector representation
of the duality group $SO(4, 20, \bR)$.

I shall now
describe one more duality with 16 supersymmetries before moving onto
theories with 8 supercharges.

\subsection{Duality of F-theory on $K3$ and Heterotic on $T^2$}

Another  interesting  application of orientifolds is in connection
with F-theory. In this subsection we concern ourselves with
F-theory compactification on  K3 to eight-dimensional
Minkowski spacetime, but these considerations are 
applicable to more general compactifications.

We have seen in $ \S{5.5}$ that to obtain an F-theory compactification,
we start with an elliptically fibered K3 
that is described by 
\bea
y^2 = x^3 + f (z) x + g (z),
\eea
where $f$ is a polynomial of  degree 8 and $g$ is a polynomial
of degree 12. 
Such a  K3 represents 
 24 stringy cosmic strings on a 2-sphere
located at the 24 points where the torus degenerates.
Typically,
the coupling constant field $\l$ will vary as we move from point to
point in the base manifold.
Consequently, there will be a non-vanishing RR background. 
Moreover, the field is allowed to undergo $ SL(2, \bZ)$ jumps.
Some of the elements of the $ SL(2, \bZ)$ like $ S$ are nonperturbative. 
Therefore, for a generic K3,
such backgrounds cannot be described  perturbatively as 
conformal field theories.

There is a special limit of  the K3 for which the modular parameter
$ \l$ of the fiber torus does not vary as we move on the sphere.
This is achieved by choosing
$f^3/g^2$ = constant.
$$
g = \phi^3 \ \ \ \ f = \alpha \phi^2 \ \ \ \ \phi (z) = {\rm
polynomial~of~degree} ~4
$$
by rescaling $y$ and $x$ we can set $\phi
= \prod^4_{a=1} (z - z_a)$. Then the $ j$ function
is given by
\bea
j(\l) &=& {4 (24\alpha)^3 \over 27 + 4 \alpha 3} = constant,
\end{eqnarray}
therefore,  $ \l$, which is the image 
under $ j^{-1}$, 
is also a constant at all points over the sphere. However
there is a nonzero $ SL(2, \bZ) $ monodromy
\bea
\label{mono}
R= \pmatrix{-1&0\cr
0&-1\cr}.
\eea
as we go around a  point  $ z =z_a$.
This is the hyperelliptic
involution of the torus which reflects both periods
of the torus without changing its modular parameter.
The discriminant is 
\begin{eqnarray}
\Delta &=& (4\alpha^3 + 27) \prod^4_{a=1} (z - z_a)^6 
\eea
which shows that the 24 7-branes are bunched in groups
of six at four points $\{ z_a\}$.
The metric on the base can be read of from \ref{metric} by
putting $ 6$ 7-branes at one point.
\bea
ds^2 = {dz d\bar z \over \prod_a |z - z_a|^{1/2} |\bar z - \bar z_a|^{1/2}}
\eea
There is a conical deficit angle of $ \pi$ at four points, otherwise
the metric is flat.
In other words, the base is $ \bT^2/\bZ_2$ and the 
fiber is $ \bT^2$ at all points except at the fixed points.

It is easy to see that this 
 K3 that is nothing but the orbifold $T^4/\bZ_2$. 
Conversely,  the K3 orbifold
$T_4/\bZ_2$ that we constructed in $ \S{5.1}$ can be viewed as an 
elliptic fibration  over $S^2$.  The K3 has coordinates
$(z_1,  z_2 )$ and the orbifold symmetry is 
$
(z_1, z_2) \sim (-z_1, -z_2).
$
It can be viewed as
\bea
\label{fibered}
{\bT^4 \over \bZ_2} = {\bT^2_{(1)} \times \bT^2_{(2)} \over \{1, R_1 R_2\}}
\eea
Let us take $ z_1$ to be the coordinate of the first
torus $ \bT^2_{1}$, and $ z_2$ 
to be the coordinate of the base.
Let $R_i$ be the operation
$ z_i \rightarrow -z_i$. 
Then the orbifold Eq.~\ref{fibered} can be viewed as
a fiber bundle with $ \bT^2_1$ as the fiber and 
$ \bT^2_{2}/\{1, R_2\}$ as the base.
The base manifold $\bT^2/\bZ_2$ is nothing but
a sphere. To see this note that  the $ \bZ_2$ symmetry
acts as  $z_2 \rightarrow -z_2$ has 
 four fixed points each with deficit angle $180^\circ = \pi$.
So the total deficit angle is $4\pi$ giving us $\int R = 4\pi$ i.e.
the correct Euler character 2 for a sphere $ S^2$.
So locally, away from the fixed points of $R_2$
the orbifold  looks like
$
\bT^2_{(1)} \times S^2 $
If we go around a fixed point of $R_2$ then the coordinate
$z_1$  of the
$\bT^2_{(1)}$ is  twisted by $R_1$, \ie, it is inverted $ (z_1\rightarrow -z_1)$,
but its modular parameter $ \l$ is unchanged.
This is precisely the   $\bZ_2$ monodromy $ R$ in Eq.~\ref{mono}.
Therefore, in this limit,  for this special configuration of  7-branes, 
the field $\l$ is constant everywhere in space and this
F-theory compactification can be described as an perturbative
orientifold \mcite{SenFThe}.
Such an identification of F-theory with an orientifold is very
useful. 

 F-theory on the K3 orbifold above is nothing but 
the orientifold
\bea
\label{orienew}
{II B~~on}~~{\bT^2 \over \{1, R I_{89} \}} \equiv 
{\bT^2 \over \{1, \O (-1)^{F_L} I_{89} \}},
\eea
if we identify $ R_1$ in Eq.~\ref{fibered} with $R= \O\FL$ and $R_2$ 
in Eq.~\ref{fibered} with $ I_{89}$.

There are $ 4$ orientifold fixed planes and $ 16$ 7-branes are required
to cancel the charge of the orientifold planes.
This fact is obvious if we
 note further that after T-dualizing  in the $ 89$ directions,
$ \O\FL I_{89}$ goes to $\O$:
\bea
 \O\FL I_{89}  = T_{89}^{-1}\O T_{89}.
\eea
Therefore, this orientifold Eq.~\ref{orienew} is T-dual
to the orientifold
\bea
{II B~~on}~~{\bT^2 \over \{1,  \O \}},
\eea
which is nothing but Type-I theory compactified on
$ \bT^2$. 
After two T-dualities the $ 32$ 9-branes turns into 7-branes.
Because of the identification $ I_{89}$ they have to move in pairs
so  effectively there are only $ 16$ of them on the orientifold.

Thus, at this special point in the moduli space, F-theory is nothing but
a T-dual of Type-I on $ \bT^2$ which in turn is dual to Heterotic on $ \bT^2$. 
We have thus established the duality 
\bea
{\rm F~theory~on~K3 } \equiv {\rm Heterotic~on}~\bT^2.
\eea

Apart from its use in understanding this duality, the orientifold
limit of F-theory has other very interesting applications.
To obtain the orientifold limit, we had to place sixteen 7-branes at
the four orientifold planes in four bunches of four. 
On the other hand, on the F-theory side, we have $ 24$ F-theory
7-branes in four bunches of $ \it six$.
What happens is that as we move the D-7branes away from
the orientifold 7-branes, then the orientifold 7-plane splits
nonperturbatively into two 7-planes \mcite{SenFThe}.
Thus, in  F-theory, the orientifold planes and the D-branes
are on an equal footing and are related nonperturbatively. 
An orientifold plane and four D7-branes  turn into
six F-theory 7-branes.
This splitting of the orientifold plane 
is very similar to the splitting of the $SU(2)$ point
into the monopole point and the dyon point 
in Seiberg-Witten theory in $ 3+1$ dimensions with gauge group
 $SU(2)$ and four quark flavors \mcite{SeWiDual}.
This similarity is not an accident but  a precise consequence
of using D3-brane probes to probe the geometry near the orientifold
plane. The worldvolume theory of D3-brane probe near an orientifold
plane and four D-7branes has  exactly the same structure
as a Seiberg-Witten theory \mcite{BDS}.


\section{Orientifolds in Six Dimensions with $ \left( 0, 1 \right)$ Supersymmetry.}

One important application of orientifolds is in the construction
of models in six dimensions with $(0, 1)$ supersymmetry
which has only $ 8$ supercharges. With only $ 8$ supercharges,
instead of $ 16$ or $ 32$,
 supersymmetry is  much less restrictive
 and therefore  
much more 
interesting dynamics is possible. At the same time,  supersymmetry
is still sufficiently restrictive to be a useful guide for checking the properties
of these theories
as well their possible duals.

The massless supermultiplets of $ (0, 1)$ supersymmetry
in terms of representations of the little group $ SU(2)\times SU(2)$
 are as follows:

1. The gravity multiplet:

\qquad a graviton ({\bf 3}, {\bf 3}), a self-dual two-form $({\bf 1}, {\bf 3})$,

\qquad  a gravitino 2({\bf 2}, {\bf 3}).

2. The vector multiplet:

\qquad a gauge boson  ({\bf 2}, {\bf 2}),

\qquad a gaugino 2({\bf 1}, {\bf 2}).

3. The tensor multiplet:

\qquad an anti-self-dual two-form $({\bf 3}, {\bf 1})$,

\qquad a fermion $2({\bf 2}, {\bf 1})$, a scalar $({\bf 1}, {\bf 1})$.

4. The hypermultiplet:

\qquad four scalars $4({\bf 1}, {\bf 1})$,

\qquad a fermion $2({\bf 2}, {\bf 1})$.

Cancellation of gravitational anomalies places restrictions on the
matter content.
Consider ${\bf V}$ vector multiplets, ${\bf H}$ hypermultiplets and
${\bf T} $ tensor multiplets. Then
the $(tr R^4)$ term in the anomaly polynomial
Eq.~\ref{anomaly} cancels only if 
\bea{
{\bf H} - {\bf V} = 273 - 29 {\bf T}.}
\eea
The $(tr R^2)^2$ term is in general nonzero, and needs to be canceled
by the Green-Schwarz mechanism.
For example, if we compactify
the heterotic string or Type-I string on a smooth K3
we obtain ${\bf T}=1$ and ${\bf H}= {\bf V} +244$.

The dynamics of $ (0, 1)$  theories in six dimensions offers many surprises
like 
the possibility of exotic phase transitions in which the
 number of (chiral) tensor multiplets changes,
or the appearance of enhanced gauge symmetry when an instanton
shrinks to zero scale size. 
Given the limitations of time it is not possible
to discuss the detailed construction in the same depth as we did in 
earlier sections. 
In this section I briefly survey two interesting phenomena:

 (i) enhanced gauge symmetry and appearance of $USp(2k)$ symmetry, and

(ii) appearance of multiple tensor multiplets,\\
and describe the utility of orientifolds
in this context. 

To organize the discussion,  a useful  starting point is the
heterotic string compactified on $ K3$. 
To begin with, in ten dimensions there are two consistent heterotic strings
that  have $N=1$ supersymmetry.

(1) The heterotic string with gauge group $SO(32)$:\\
The strong coupling limit of this theory is Type-I string.
We have already used this duality in $ \S {6}$.  

(2) The heterotic string with gauge group $ E_8\times E_8$:\\
The strong coupling limit of this theory is M-theory
on $ \bM^{10} \times ({\bf S^1}/\bZ_2)$ \mcite{HoWi}.  The generator of $ \bZ_2$
reflects the coordinate of the circle so the resulting
orbifold $ ({\bf S^1}/\bZ_2)$ is nothing but a line segment
${\bf I}:  [0, \pi R]$. $ \bM^{10}$ is ten-dimensional Minkowski spacetime.
The two $ E_8$ factors live on the two
`end of the world' boundaries of this manifold. Gravity lives in the bulk
and on the boundary. In the bulk we have eleven dimensional
supergravity which corresponds to $ N=2$ Type-IIA supergravity
in ten dimensions upon dimensional reduction on  a circle. Boundary conditions
break it to $ N=1$ supergravity of heterotic string.

The supergravity sector of the heterotic string  is identical
to the Type-I theory. It
contains, along with a graviton,  a dilaton, and the fermions,
a 2-form field $B$ with field strength  $ H$
which satisfies the Bianchi identity
\bea
\label{bianchi}
dH= {\it tr } (R\wedge R) - {\it tr }(F\wedge F),
\eea
where $ R$ is the curvature 2-form and $ F$ is the gauge field strength
2-form.
Integrating the Bianchi identity gives the constraint that the
integral of the right hand side of Eq.~\ref{bianchi} should
vanish on a manifold without a boundary. In particular,
when we compactify
the string  on K3, the integral of the right hand side over K3 should vanish.
Now $ \int_{K3} {\it tr } (R\wedge R)$ is the Euler character of K3 which equals
$ 24$. Therefore, the integral $ \int_{K3}  {\it tr }(F\wedge F)$, which
is the Pontryagin index or the instanton number of the gauge field
on K3, should also equal $ 24$.
A consistent heterotic compactification on K3 is possible only if 
the gauge field is also nontrivial such that there
are $ 24$ instantons on K3.
An instanton on K3 looks like a solitonic 5-brane
\mcite{Stro} in ten dimensions that
fills flat six-dimensional Minkowski spacetime.

The heterotic compactification can become singular
in certain limits. One singularity that will concern us here
is when  the scale size of an instanton shrinks to
zero. 
When a 5-brane instanton shrinks, the geometry in the transverse
space becomes singular and it develops a long throat.  
Deep down the throat the coupling constant grows and
the perturbative description of the 5-brane breaks down. 
We would like understand what happens near such a singularity.

{}From our experience with the moduli space of supersymmetric
gauge theories and string theories,
typically a singularity in the  moduli space
indicates that additional
states are becoming massless at that point in the
moduli space. The singularity occurs because 
we have incorrectly  integrated out these states 
that are  massless. We expect that 
the singularities in the instanton moduli space also
has a similar physical explanation. 
The small instanton singularity  has
a completely different  physical resolutions in the two heterotic strings,
each remarkable in its own way.

(i) Small instantons in $SO(32)$ heterotic string:\\
If there are $ k$ small instantons that coincide, then there
are additional massless gauge bosons which give
 rise to an additional
$ USp(2k)$ gauge symmetry.
In heterotic theory, this remarkable conclusion was arrived at from 
considerations of the  ADHM
construction of instantons\mcite{WittSmal}. 

This phenomenon, which cannot be described
in  a weakly coupled conformal field theory in heterotic compactifications,
has a simple perturbative description in terms of a Dirichlet 5-brane
in the dual Type-I orientifold theory.
In particular, the enhanced $USp(2k)$ symmetry when $k$ small
instantons coincide can be understood in terms of coincident
5-branes with a specific symplectic projection in the open string
sector that is determined by the consistency of the world-sheet
theory.

(ii) Small  instantons in $E_8 \times E_8$ heterotic string:\\
These are even more unusual to understand. The picture is clearer in
 the dual M-theory.
Consider the dual compactification of M-theory on 
K3 $ \times(\bS^1/\bZ_2)$. The gauge fields and the 
corresponding instantons
in the two $ E_8$ factors are confined to the two boundaries.
M-theory contains a solitonic 5-brane which carries unit 
charge exactly like the heterotic
instanton 5-brane. When one of the instanton 5-brane in the
boundary $ E_8$ gauge theory shrinks to zero size,
the resulting singularity corresponds to an M-theory
solitonic 5-brane stuck to  the boundary.
One of the possibilities consistent with the Bianchi identity of
$ H$ and anomaly
cancellation is that the M-theory 5-brane can be
emitted from the `end of the world'  into into the bulk \mcite{SeWiComm}.  
The worldvolume theory of an M-theory 5-brane
contains  a tensor multiplet in its worldvolume theory \mcite{CHS}. 
The 5-brane fills the noncompact six-dimensional space-time $ \bM^6$.
Therefore, in the process of emission of an M-theory 5-brane
into the bulk, the number of tensor multiplets in the six-dimensional
theory can increase
by one.

Multiple tensor multiplets are not possible with
usual Calabi-Yau
compactifications.
However, as we shall see in $ \S{7.2}$,
one can easily construct orientifolds that have this property.

\subsection{Symplectic Gauge Groups}

Let us consider Type-IIB theory on a  $ K3$ orbifold  $\bT^4/\bZ_2$
and orientifold  this theory further by $ \{1,\O\}$ to obtain
Type-I theory  on
the $ K3$ orbifold. Details of this orientifold  
can be found in \mcite{GiPo}. Here we shall point out
some of the salient features.

(a) The orientifold group is 
\bea
G = \{1, I_{6789}, \Omega, \Omega I_{6789}\}.
\eea

(b) Fixed planes of $ \Omega$ are orientifold 9-planes filling
all space,
which are identical to those in Type-I theory in ten dimensions
discussed in $ \S{4}$. 
The charge of the 
orientifold plane is $ -32$ as before requiring $ 32$ D9-branes
to cancel it. Fixed planes of $ I_{6789} \O$ are orientifold  5-planes
located 
at the $ 16$ fixed planes of $ I_{6789}$
each of charge $ -2$. The net charge is again
$ -32$ requiring $ 32$ D5-branes.

(c) There are now four open string sector: 55, 59, 95, 99,
depending on what type of brane the two ends
of an open string end on.

(d) A D-5 brane has the
same charge as a small instanton and is  dual
of the solitonic 5-brane in the
heterotic string. 

(e) Tadpole cancellation determines  the matrix $\gamma_{\O, 9}$,
which implements the $ \O$ projection in the 99 sector, to be  symmetric
as in $ \S{4}$.
Therefore, the  gauge group of the $ 32$ 9-branes is an
orthogonal subgroup of $U(32)$ after the $ \Omega$ projection.
This group is further broken by the $I_{6789}$ projection.

(f) Consistency requirements determine
that the matrix $ \gamma_{\O, 5}$, 
which implements the $ \O$ projection in the 55 sector,
must be  antisymmetric if $ \g_{\O, 9}$ is symmetric.
This follows from a somewhat subtle argument by \mcite{GiPo}
that involves considerations of  factorization and the action
of $ \O^2$ in the $ 59$ sector. We do not repeat the argument
here and refer the reader to \mcite{GiPo} for details.
When $2k$ 5-branes coincide, we get a symplectic
subgroup $ USp(2k)$ of $U(2k)$ after the $ \O$ projection,
by the arguments discussed at the end of $ \S{4.5}$.
Thus, the small instantons and the enhanced $ USp(2k)$ gauge
symmetry have a very
simple perturbative description in terms of  D 5-branes in Type I. 

There are many interesting aspects of this model which have  been
analyzed in great detail in \mcite{Six}. 
A nonperturbative description using F-theory can be found
in \mcite{SenAnon}.

\subsection{Multiple tensor Multiplets}

We now describe models with $ (0, 1)$ supersymmetry
with multiple tensor multiplets. With conventional
Calabi-Yau compactifications,  the only
way to obtain $(0, 1)$ supersymmetry is to compactify
either the Heterotic or the Type-I theory on a $ K3$.
These compactifications give only a single tensor
multiplet.

For Type-II strings, compactification on a $K3$ leads
to $N=2$ supersymmetry as we saw in $ \S{5.3}$. 
One cannot obtain
lower supersymmetry with  Calabi-Yau compactification.
One way to reduce supersymmetry further is to take an orientifold  so that
only one combination of the left-moving and the right-moving supercharges
that is preserved by the orientation-reversal survives. 
If we wish to obtain a large number of tensor multiplets, a
natural starting point for orientifolding is
the Type-IIB theory compactified on $K3$
which has 21 tensor multiplets of $ (0, 2)$ supersymmetry.
A tensor multiplet of $(0, 2)$ supersymmetry is a sum 
 of a  tensor-multiplet ({\bf T}) and a hyper-multiplet  ({\bf H}) of
$ (0, 1)$ supersymmetry.  

If we use the projection  $(1+\Omega)/2$ then we get Type I
theory on K3.
Under $\Omega$, the 4-form $D_{i j k l}$ and the 2-form $B_{i j}$  are odd. 
Therefore all zero modes of these fields are also projected out.
Only the zero modes of the field $ B'_{ij}$ survive which gives
one self-dual and one anti-self-dual tensor in six dimensions.
The self-dual tensor is required in the gravity multiplet.
So, we end up with a single tensor multiplet.

This counting suggests a generalization.
If the K3 orbifold 
has  a  $ Z_2$ symmetry  with generator
$ S$, then we can consider taking $ \{ 1, \O S\}$ as the orientifold
group. If we want $ N=1$ supersymmetry, then the requirement
is that $ S$ should not break supersymmetry further.
This is ensured if it leaves the holomorphic two-form of the
$ K3$ invariant.
But, $ S$ can  have nontrivial action
on other harmonic forms of  the $ K3$. 
Some of the zero modes of $ D_{ijkl}$,
can be even with respect to 
the combined action of $ \O S$ 
even if they are odd with respect to $ \O$ alone.
These can give rise to additional tensor multiplets that we
are interested in.

Concretely, let us consider an example of such a symmetry
discussed in \mcite{DaPaI}.
Consider a K3 orbifold $ {\bf T}^4/\bZ_2$ that we have been
discussing in $ \S{5}$. Such a K3
 admits a $ \bZ_2$ involution
\begin{eqnarray}
S &:& (z_1, z_2) \rightarrow (-z_1 + {1 \over 2}, - z_2 + {1 \over 2}).
\eea
$ S$ leaves the holomorphic 2-form $ dz^1\wedge dz^2$ of the
K3 orbifold, and in fact all forms coming from the untwisted sector
of the orbifold invariant. Let us look at its action on the twisted sector.
It is easy to see from Figure~\ref{square} that 
$S$ takes the 16 fixed points of $I_{6789}$ into each other so out
of the $16_a$ anti-self-dual 2-forms coming from the twisted sector, 
eight are odd and eight are even.

To obtain an anti-self-dual 2-form in six dimensions as a zero mode
of the 4-form in ten dimensions, we use separation of variables
to write the 4-form as
$ D^{(4) }= B^{\left( 2 \right)}_{\a}\wedge f^2_{\alpha}, \a=1,\ldots, 19$ where 
$ f^2_{\alpha}$ is one of the anti-self-dual harmonic 2-forms
on K3 which depends only on the coordinates
of $ K3$ and  $B^{\left( 2 \right)}_{\a} $ depends only on the
non-compact coordinates. Because
$ f^2_{\alpha}$ is harmonic,  $B^{\left( 2 \right)}_{\a}$ is
a massless field in six dimensions.
By the self-duality of $D^{(4)}$ in ten dimensions, 
$ B^{\left( 2 \right)}_{\a}$ is anti-self-dual in the six Minkowski dimensions.
Now, if we use the combined 
projection, ${(1 + \Omega S) \over 2}$ instead of ${(1 + \Omega) \over
2}$ then eight tensors coming from the eight  $ f^2_{\alpha}$'s that are
odd under $ S$ survive and the remaining are projected out.
In addition there is one more tensor multiplet that comes from the
zero mode of $ B'_{ij}$ as in Type-I.
Altogether, we get $ \bT=9$.

The orientifold group is
\bea
G=\{ 1,  I_{6789}, \O S, \O I_{6789}S\}.
\eea
We shall not discuss the open string sector here but it can be
found in \mcite{DaPaI}. One interesting aspect of models
with multiple tensors  is worth mentioning. 
The cancellation
of gauge
and gravitational anomalies in these models requires
an extension of of the Green-Schwarz mechanism found by
Sagnotti \mcite{SagnII} in which more than
one tensors participate in the  anomaly cancellation.
 Details of anomaly cancellation for the model
described above can be found in \mcite{DaPaI}).

The model has an M-theory dual \mcite{SenMThe} that makes
use of the observation that Type-IIB on $ K3$ is dual to
M-theory on $ \bT^5/\bZ_2$ \mcite{DaMu,WittFive}.

There are a number of other ways to obtain multiple tensors
in string models.
Orientifolds of K3 orbifolds where the orbifold group is other than
$ \bZ_2$ typically give multiple tensors \mcite{GiJoK3Or,DaPaII,GiJoMult}.
Yet another interesting variation is to accompany the action of $ \O$
with additional phases in the twisted sectors of the $ \bZ_2$ orbifold
symmetry \mcite{PolcTens,BlZa,DaPaIII,GoMu}. This is the analog of 
discrete torsion for $ \bZ_N\times \bZ_N$ orbifolds \mcite{VafaModu}.

\section*{Acknowledgments}

I would like to thank the organizers of the ICTP Summer school for inviting
me deliver these lectures, and Rajan Pawar and M.~R.~Shinde for help
with figures and 
with TeXing the manuscript.

\end{document}